\documentclass[preprint2]{aastex6}
\usepackage[fleqn]{amsmath}
\usepackage{amssymb}

\usepackage[normalem]{ulem}

\shortauthors{J. Pe\~na et al.}

\begin{document}


\title{Asteroids' Size Distribution and Colors from HiTS}
%lessons for wide field surveys
%with Ecliptic Latitude}


\author{J. Pe\~na\altaffilmark{1,2},
  C. Fuentes\altaffilmark{1,2},
  F. F\"orster\altaffilmark{3,2},
  J. Mart\'inez-Palomera\altaffilmark{4,1,3},
  G. Cabrera-Vives\altaffilmark{5,2}
  J.C. Maureira\altaffilmark{3},
  P. Huijse\altaffilmark{6, 2},
  P.A. Est\'evez\altaffilmark{7,2},
  L. Galbany\altaffilmark{8},
  S. Gonz\'alez-Gait\'an\altaffilmark{9, 3, 2},
  Th. de Jaeger\altaffilmark{4, 1, 2}
}
\email{jpena@das.uchile.cl}


\altaffiltext{1}{Departamento de Astronom\'ia, Universidad de Chile, Camino del Observatorio 1515, Las Condes, Santiago, Chile.}
\altaffiltext{2}{Millennium Institute of Astrophysics, Chile.}
\altaffiltext{3}{Center for Mathematical Modeling, Beaucheff 851, 7th floor, Santiago, Chile.}
\altaffiltext{4}{Department of Astronomy, University of California, Berkeley, CA 94720-3411, USA.}
\altaffiltext{5}{Department of Computer Science, Universidad de Concepci\'on, Chile.}
\altaffiltext{6}{Informatics Institute, Universidad Austral de Chile, Chile.}
\altaffiltext{7}{Electrical Engineering Department, University of Chile, Chile.}
\altaffiltext{8}{Departamento de F\'isica Te\'orica y del Cosmos, Universidad de Granada, E-18071 Granada, Spain.}
\altaffiltext{9}{CENTRA, Instituto Superior T\'ecnico, Universidade de Lisboa, Portugal.}


\begin{abstract}
%\jpz{The abstract should summarize concisely the content and conclusions of the article. The abstract should be a single paragraph of not more than 250 words.}

We report the observations of solar system objects during the 2015 campaign of the High cadence Transient Survey (HiTS). We found 5740 bodies (mostly Main Belt asteroids), 1203 of which were detected in different nights and in $g'$ and $r'$.
%in more than 1 night (allowing to constrained their orbits).
%OurTo find these bodies we used a linking algorithm that takes into account Earth's motion and assumes circular orbits.
Objects were linked in the barycenter system and their orbital parameters were computed assuming Keplerian motion.
%Using known asteroids in this survey, we find that our typical errors in semi-major axis, eccentricity, inclination, barycentric distance and observer's distance are \rv{$\sigma_a\sim 0.05 ~{\rm au}$}, \rv{$\sigma_{\rm e} \sim 0.06 $}, \rv{$\sigma_i\sim 0.5 ~{\rm deg}$}, \rv{$\sigma_r\sim 0.12 ~{\rm au}$} and \rv{$\sigma_{\Delta}\sim 0.12 ~{\rm au}$}, respectively. 
We identified 6 near Earth objects, 1738 Main Belt asteroids and 4 Trans-Neptunian objects.
We did not find a $g'-r'$ color--size correlation for $14<H_{g'}<18$ ($1<D<10$ km) asteroids. We show asteroids' colors are disturbed by HiTS' 1.6 hour cadence and
%does not allow to compute asteroid's color properly. 
estimate that observations should be separated by at most %Colors should be measured in less than 
14 minutes to avoid confusion in future wide-field surveys like LSST. %s is a very important constrained for future surveys such as for LSST.
The size distribution for the Main Belt objects can be characterized as a simple power law with slope $\sim0.9$ %for the 2015 campaign
, steeper than in any other survey, while
%The size distribution 
data from HiTS 2014's campaign is consistent with previous ones (slopes $\sim0.68$ at the bright end and $\sim0.34$ at the faint end). This difference is likely due to the ecliptic distribution of the Main Belt since 2015's campaign surveyed farther from the ecliptic than did 2014's and most previous surveys.

\end{abstract}



\section{Introduction}

Our solar system (SS) is currently understood to have emerged from a protoplanetary disk (\citealt{Armitage.2017} and references therein). While planets grew and migrated (see \citealt{Horner.2013} for a summary), thousands of planetesimals formed, grew bigger and broke into smaller pieces in a process that can still be studied in the stable reservoirs of minor bodies, namely the Main Belt (MB), Jovian Trojans, Neptunian Trojans, and Trans-Neptunian Objects (TNOs) \citep{Sheppard.2006}. This evolution has left its mark in the orbital distribution  of the Main Belt and in its size distribution (SD).
%, while planet migration sculpted it,
%creating gaps in semi-major axis, increasing inclinations, and in the evolution of the size distribution (SD).

The SD provides a direct glimpse into the collisional history of minor bodies. Under collisional equilibrium the SD is described by a power law $N(H)\propto 10^{\alpha H}$ and $N(D)\propto D^{-q}$ ($H$ the absolute magnitude and $D$ the body's diameter, with $q=5\alpha+1$) with $q=3.5$ and $\alpha=0.5$ \citep{Dohnanyi.1969}. In Table \ref{tab:slopes} we provide a summary of Main Belt surveys, including SD best fit and filter information. Having multiple filters provides asteroids' surface colors that might be related to composition and collisional history.

%Various studies in different surveys and data sets have found different values for $\alpha$ and $q$. 
\cite{Ivezic.2001} analyzed Sloan Digital Sky Survey (SDSS) data \citep{York.2000} and separated the MB by color, finding $\alpha \sim 0.61\pm0.01$ for brighter bodies while they found $\alpha=0.24\pm0.01$ %($q=2.20\pm0.05$)
for \textit{``red''} bodies, $\alpha=0.28\pm0.01$ %($q=2.40\pm0.05$) 
for \textit{``blue''} bodies and $0.25\pm0.01$ %($q=2.3$) 
for \textit{``blue''} and \textit{``red''} combined (\textit{``red''} and \textit{``blue''} by their definition, associated to S-type and C-type respectively).
For S-type and C-type asteroids, they found both their SDs to have a break at $D\sim5$ km, and attributed this feature to a color-size dependence for bodies smaller than 5~km.
\cite{Parker.2008}, using a more updated data set (an early version of the 4th release of the SDSS Moving Object Catalog, \citealt{Ivezic.2010}) divided the MB by semi-major axis $a$ and analyzed the absolute magnitude $H$, finding similar slopes for bright bodies but steeper slopes for smaller objects ($\alpha\sim0.42$). In both cases they found the SD gets flatter with $a$. \cite{Parker.2008} also analyzed individual asteroid families, finding slopes $\alpha$ varying from 0.37 to 1.04 at the bright end and $\alpha$ from 0.1 to 0.62 at the faint end.
\cite{Yoshida.2003} analyzed the size distribution for $\sim1000$ small asteroids from SMBAS (Subaru Main Belt Asteroid Survey). Observations were only separated by $\sim$2 hours, affording only rough distance estimates. They report no break with a single power slope $q=2.19\pm0.02$ ($\alpha=0.24$) for the entire MB SD (for $D>0.5$ km) which is similar to the one found by \cite{Ivezic.2001} in SDSS. They also found that the SD gets flatter with $a$. %and also for MB zones: $q=2.37\pm0.003$ ($\alpha=0.47$) until $D>0.23$ km for the Inner Belt ($2.0<a<2.6$ au); $q=2.15\pm0.003$ ($\alpha=0.43$) until $D>0.34$ km for the Middle Belt ($2.6<a<3.0$ au); and $q=1.98\pm0.003$ ($\alpha=0.4$) until $D>0.49$ km for the Outer Belt ($2.6<a<3.0$ au), finding that the slope get flatter with distance 

%\citealt{Parker.2008} but not as steep).

\cite{Yoshida.2007} using a new set of data from SMBAS measured $B-R$ and $H$, getting broken power laws for small bodies: $q=2.29\pm0.02$ ($\alpha=0.26$, similar to faint bodies from \citealt{Ivezic.2001}) for small objects ($D<1$ km) and $q=2.75\pm0.02$ ($\alpha=0.35$) for bright objects (between the values from \citealt{Ivezic.2001} and \citealt{Parker.2008}); they also separated the bodies by color (based on a slight low density in their color histograms), finding that S-like (redder) bodies have $q=2.29\pm0.02$ ($\alpha=0.26$) at the faint end ($D<1$ km) and $q=3.44\pm0.09$ ($\alpha=0.49$) at the bright end, while C-like (bluer) bodies could be characterized by a single slope of $q=2.33\pm0.03$ ($\alpha=0.27$). \cite{Lin.2015} analyzed 150 asteroids finding an SD compatible with the slopes found by \cite{Yoshida.2007}; they also found that S-like bodies are more common in the inner region of the MB while C-like bodies dominate the region beyond 2.82 au of the MB.
\cite{Wiegert.2007} analyzed 1525 MB bodies with an arc of $\leqslant2$ days measured in $g'$ or in $r'$ with the Canada-France-Hawaii Telescope (CFHT); they found a very clear difference of slopes between $g'$ %($q=1.87\pm0.05$)
and $r'$ %($q=2.44\pm0.07$)
and a varying slope with distance, getting a steeper slope between $2.6<a<3.0$ au (parameter obtained assuming circular orbits); these results were discarded by \cite{August.2013} because they used data beyond the limiting magnitude. 
\cite{August.2013} used $\sim17,000$ MB bodies from CFHT Legacy Survey (CFHTLS) with measurements in $g'$ and $r'$, getting $\alpha=0.39\pm0.01$ in $g'$ and $r'$ for all bodies %and a slightly flatter slope for outer bodies. %while when separating by distance, they get $\alpha\approx0.39\pm0.02$ in $g'$ and $r'$ between $2.0<a<3.0$ au and $\alpha\approx0.36\pm0.005$ in $g'$ and $r'$ between $3.0<a<4.0$ au; 
without finding a color-size dependence, but they do recover a flatter slope at higher distance (all of this between $15<H<17$). \cite{August.2013} suggest this slope-distance dependence is produced by a difference in composition, in that the inner MB is dominated by S--type bodies and the outer MB by C--type, although they do mention that it is not clear how this differentiation affects the slope and they do not do a color analysis such as that of \cite{Ivezic.2001} and \cite{Yoshida.2007}.
One year earlier, \cite{Gladman.2009} analyzed $\sim1000$ small bodies with time ranges of more than three nights (in the Sub-kilometer asteroid diameter survey, SKADS), allowing a good calculation of $H$, finding $\alpha=0.38$ (in between the ones found by \citealt{Ivezic.2001} and \citealt{Parker.2008}, but very similar to the one by \citealt{August.2013}). SKADS also has color measurements, but \cite{Gladman.2009} did not find any bimodality as in SDSS \citep{Ivezic.2001} or as claimed in \cite{Yoshida.2007}.
\cite{Masiero.2011} computed asteroid diameters from \textit{Wide-field Infrared Survey Explorer} (WISE) data \citep{Wright.2010} and found that the SD follows a slope similar to the one find by \cite{Gladman.2008} for small bodies.
\cite{Ryan.2015} analyzed $\lesssim2000$ from \textit{Spitzer}'s MIPSGAL and Taurus surveys \citep{Carey.2009, Rebull.2010}, obtaining $q=3.34\pm0.05$ ($\alpha=0.47\pm0.01$) for MB bodies between 2 and 25 km (which seems an in-between value from bright and faint slopes from previous surveys such as SDSS), and different slope values when separating by taxonomic type, 
%they find $q=3.61\pm0.04$ ($\alpha=0.52\pm0.01$) for C-type and $q=2.91\pm0.11$ ($\alpha=0.38\pm0.02$) for S-type (both with $5<D<25$ km), 
although it seems they do not take into consideration their completeness limit (of 6.65 km or 15.75 in $H$ according to them).
In summary, there is a wide range of values for the $\alpha$ parameter, especially for faint bodies ($D\lesssim5$ km). This is probably caused by differences in data reduction, orbital parameter determination and limiting magnitudes of each work. The most complete data set comes from SDSS (which uses known bodies for their orbital parameters), getting $\alpha\sim0.61$ for bright bodies. For faint bodies, the values vary from 0.25 \citep{Ivezic.2001} to 0.38 by \cite{Gladman.2009} (which is the largest survey published with good orbital parameter estimations) to $\sim0.42$ by \cite{Parker.2008} (again in SDSS). Many of these surveys have found that the SDs get flatter with $a$, while only some of these studies have found a slight color-size dependence.

In this paper we show our results finding asteroids in the 2015A campaign of the High cadence Transient Survey (HiTS). A fraction of our data ($\sim$1,700) have measured arcs of $\sim$24 hours, allowing acceptable orbital solution for $H$ analysis. We also have $g'-r'$ for $\sim$1,200 of them, allowing some color analysis. In section \ref{sec:data}, we present the HiTS data. In section \ref{sec:analysis} we explain our detection linking algorithm to get the different asteroids. In section \ref{sec:results} we show our results: orbital parameters distribution, apparent and absolute magnitude distribution and color analysis (mainly for Main Belt objects). Finally, in section \ref{sec:conc} we present our conclusions for the 2015A campaign and we put them in contrast to results from 2014A campaign.

\begin{deluxetable*}{lclcCCCC}%[b!]
\tablecaption{Summary of SD slopes from different surveys\label{tab:slopes}}
\tablecolumns{8}
%\tablenum{1}
\tablewidth{0pt}
\tablehead{
\colhead{Survey} &
\colhead{Date\tablenotemark{a}} &
\colhead{$m_{lim}$} &
\colhead{Population} &
\colhead{Criterion} &
\colhead{N\tablenotemark{b}} &
\colhead{$\alpha$\tablenotemark{c}} &
\colhead{Size Range} %\\
}
\startdata
SDSS\tablenotemark{d} & 2001 & $r*<21.5$ & MB & $1.5\lesssim a\lesssim4$ \text{au} & 670,000 & $0.61\pm.01$ & $D>5$ \text{km} \quad ($H\lesssim 15.7$) \\
 & & & & & & $0.25\pm.01$ & $D<5$ \text{km} \quad ($H\gtrsim 15.7$) \\
 & & & \textit{``blue''} & $a^*<0$ & 467,000 & $0.61\pm.01$ & $D>5$ \text{km} \\
 & & & & & & $0.28\pm.01$ & $D<5$ \text{km} \\
 & & & \textit{``red''} & $a^*>0$ & 203,000 & $0.61\pm.01$ & $D>5$ \text{km} \\
 & & & & & & $0.24\pm.01$ & $D<5$ \text{km} \\
\hline
SMBS\tablenotemark{e} & 2003 & $R<24.4$ & MB & 2<a<3.5 \text{au} & $\sim500$ & $0.238\pm.004$ & $.5<D<1$ \text{km} \quad ($18.3<H_R<19.8$) \\
 & & & inner MB & 2<a<2.6 \text{au} & $\sim200$\tablenotemark{f} & $0.274\pm.006$ & $.23<D<1$ \text{km} \quad ($18.3<H_R<21.4$) \\
 & & & middle MB & 2.6<a<3.0 \text{au} & $\sim250$\tablenotemark{f} & $0.230\pm.006$ & $.34<D<1$ \text{km} \quad ($18.3<H_R<20.6$) \\
 & & & outer MB & $3.0<a<3.5$ au & $\sim50$\tablenotemark{f} & $0.196\pm.006$ & $.49<D<1$ \text{km} \quad ($18.3<H_R<19.8$) \\
\hline
SMBS\tablenotemark{g} & 2007 & $R<25$ & MB & 2<a<3.5 \text{au} & $\sim800$ & $0.258\pm.004$ & $D<1$ \text{km} \quad ($17.8<H<20.2$) \\
 & & & & & $\sim200$ & $0.350\pm.004$ & $D>1$ \text{km} \quad ($14.6<H<17.4$) \\
 & & & S--like & $B-R>1.1$ & -- & $0.058\pm.004$ & $0.3<D<1$ \text{km} \quad ($17.4<H<20.2$) \\
 & & & & & -- & $0.488\pm.018$ & $D>1$ \text{km} \quad ($15.4<H<17.0$) \\
 & & & C--like & $B-R<1.1$ & -- & $0.266\pm.006$ & $D>0.6$ \text{km} \quad ($14.6<H<20.2$) \\
\hline
CFHTLS\tablenotemark{h} & 2007 & $g'<22.5$ & MB & 2.0<a<3.5 \text{au} & 185 & 0.37\pm0.01 & 0.6<D<10 \text{km} \\
 & & & inner MB & 2.0<a<2.6 \text{au} & 77 & 0.316\pm0.012 & 0.6<D<4 \text{km} \\
 & & & middle MB & 2.6<a<3.0 \text{au} & 79 & 0.370\pm0.012 & 0.8<D<6.3 \text{km} \\
 & & & outer MB & 3.0<a<3.5 \text{au} & 29 & 0.320\pm0.014 &  1<D<6.3 \text{km}\\
 &  & $r'<21.75$ & MB & 2.0<a<3.5 \text{au} & 423 & 0.488\pm0.014 & 1<D<10 \text{km} \\
 &  & & inner MB & 2.0<a<2.6 \text{au} & 238 & 0.40\pm0.01 & 1<D<7.9 \text{km} \\
 &  & & middle MB & 2.6<a<3.0 \text{au} & 143 & 0.478\pm0.014 & 1.3<D<7.9 \text{km} \\
 & & & outer MB & 3.0<a<3.5 \text{au} & 42 & 0.45\pm0.016 &  1.6<D<6.3 \text{km}\\
\hline
SDSS\tablenotemark{i} & 2008 & $r'<21.5$ & inner MB & 2.0<a<2.5 \text{au} & 30,702 & $0.76$ & H<14 \quad (D>7 \text{km})\tablenotemark{*} \\
 & & & & & & $0.46$ & H>14 \\
 & & & middle MB & 2.5<a<2.82 \text{au} & 32,500 & $0.73$ & H<13.5 \quad (D>9 \text{km})\tablenotemark{*} \\
 & & & & & & 0.42 & H>13.5 \\
 & & & outer MB & 2.82<a<3.6 \text{au} & 24,367 & 0.56 & H<13.5 \\
 & & & & & & 0.4 & H>13.5 \\
\hline
SKADS\tablenotemark{j} & 2009 & $R<23.5$ & MB & 2.0<a<4.0 \text{au} & \sim1000 & 0.38 & 14.8<H_R<17.4 \quad (1<D<5\text{km})\tablenotemark{**} \\
\hline
CFHTLS\tablenotemark{k} & 2013 & $g'\lesssim23$ & MB & 2.0<a<4.0 \text{au} & 7285 & 0.39\pm0.01 & 15<H_{g'}<17.6 \quad (1<D<5\text{km}) \\
 & & & inner MB & 2.0<a<3.0 \text{au} & \text{--} & 0.39\pm0.02 & 15<H_{g'}<17.6 \\
 & & & outer MB & 3.0<a<4.0 \text{au} & \text{--} & 0.35\pm0.01 & 15<H_{g'}<17.6 \\
 & & $r'<22.5$ & MB & 2.0<a<4.0 \text{au} & 9671 & 0.39\pm0.01 & 15<H_{g'}<17.1 \\
 & & & inner MB & 2.0<a<3.0 \text{au} & \text{--} & 0.39\pm0.01 & 15<H_{g'}<17.1 \\
 & & & outer MB & 3.0<a<4.0 \text{au} & \text{--} & 0.365\pm0.004 & 15<H_{g'}<17.1 \\
 \hline
\textit{Spitzer}'s\tablenotemark{l} & 2015 & -- & MB & 2.06<a<3.65 \text{au} & 1865 & 0.47\pm0.01 &  2<D<25\text{km} \\
 & & & C-Type & p_V<0.08 & $\sim600$ & 0.52\pm0.01 &  5<D<25 \text{km} \\
 & & & S-Type & 0.15<p_V<0.35 & $\sim400$ & 0.38\pm0.02 &  5<D<25 \text{km} \\
 \hline
HiTS 2014 & \small{this work} & $g'<22.5$ & MB & $1.3<a<4.2$ \text{au} & 1,729 & 0.68^{+0.17}_{-0.09} & 11<H_{g'}<14 \quad (10\lesssim D\lesssim30\text{km}) \\
 & & & & & & 0.34^{+0.04}_{-0.11} & 14<H_{g'}<17 \quad (1\lesssim D\lesssim10\text{km}) \\
HiTS 2015 & \small{this work} & $g'<22$ & MB & $1.3<a<4.2$ \text{au} & 129 & 0.88^{+0.09}_{-0.08} & 14<H_{g'}<16.5  \quad (1\lesssim D\lesssim10\text{km})\\
\enddata
\tablenotetext{a}{Publication date of the study.}
\tablenotetext{b}{Number of bodies used in the analysis.}
\tablenotetext{c}{Slope of the SD using $H$}
\tablenotetext{d}{\cite{Ivezic.2001}}
\tablenotetext{e}{\cite{Yoshida.2003}}
\tablenotetext{f}{From Figure 11 in \cite{Yoshida.2003}}
\tablenotetext{g}{\cite{Yoshida.2007}}
\tablenotetext{h}{\cite{Wiegert.2007}}
\tablenotetext{i}{\cite{Parker.2008}}
\tablenotetext{j}{\cite{Gladman.2009}}
\tablenotetext{k}{\cite{August.2013}}
\tablenotetext{l}{\cite{Ryan.2015}}
\tablenotetext{*}{$D$ calculated using albedo $p_V=0.1$}
\tablenotetext{**}{$D$ calculated using albedo $p_V=0.1$ and an average color of $V-R\simeq0.4$ (see Figure 15 in \cite{Gladman.2009}}
\end{deluxetable*}

\section{Data} \label{sec:data}

\subsection{HiTS observations}
HiTS was a survey aimed to discover and follow up transients, especially the earliest hours of supernova explosions. For this, it combined high cadence with a high limiting magnitude and a wide field of view. These characteristics offer the opportunity to do science in various topics other than supernovae \citep{2016ApJ...832..155F, Forster.2018} such as RR Lyrae \citep{Medina.2017, Medina.2018}, SS minor bodies (\citealt{Pena.2018} and this work) and automatic classification of variable sources \citep{Martinez.2018}.

HiTS observations were obtained with the Dark Energy Camera (DECam) mounted at the prime focus of the Blanco 4m telescope at the Cerro-Tololo
International Observatory. DECam covers a 3 square degree field of view with a mosaic of $\sim60$ ccd of 2Kx4K pixels, yielding a  $0.27\arcsec$/pixel resolution \citep{DECAM}.

\begin{figure}
   \centering
   \includegraphics[width=\hsize]{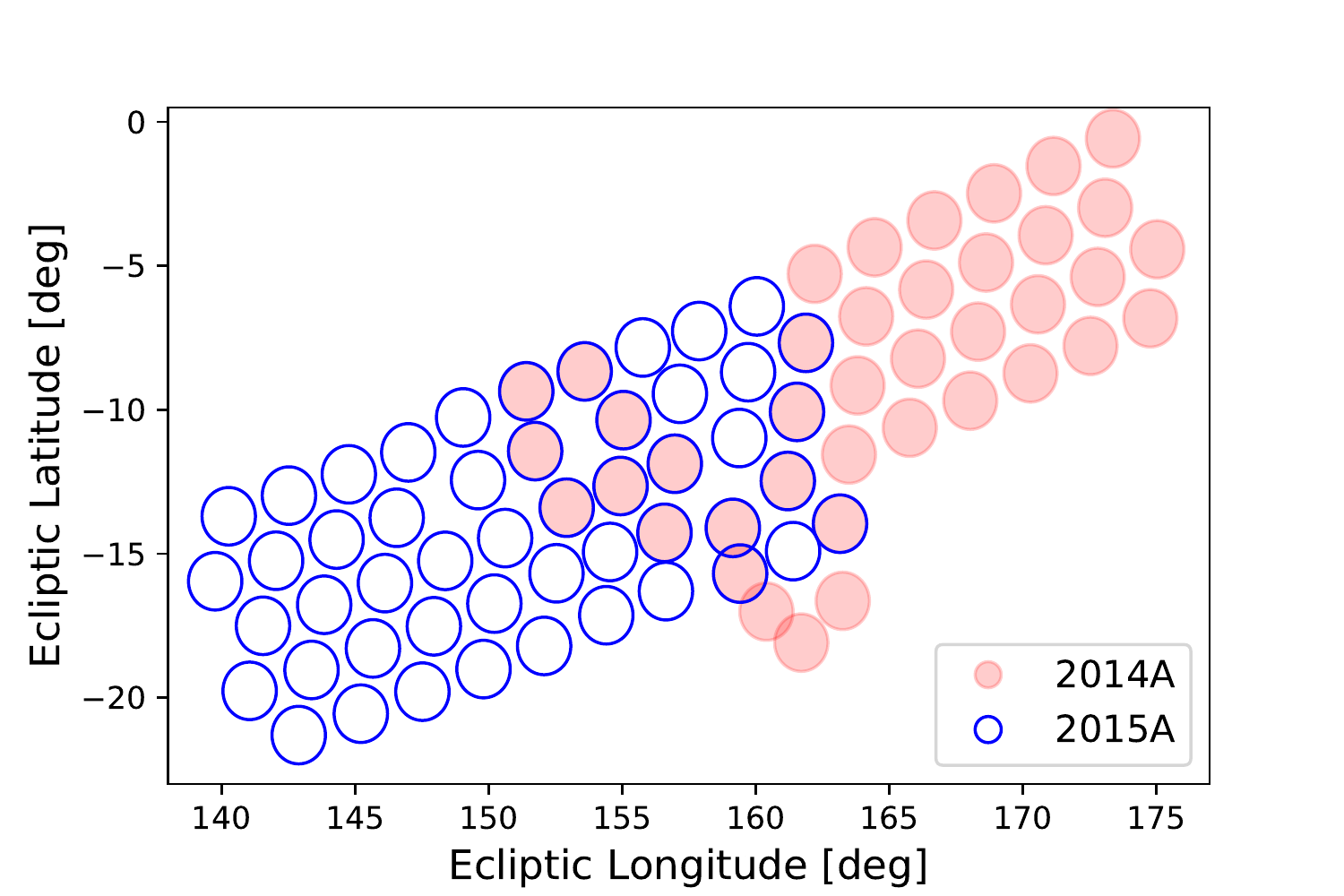}
   \caption{HiTS pointings for the 2014A and 2015A campaigns in geocentric ecliptic coordinates.}
   \label{fig:ecliptic}%
\end{figure}

HiTS was run in three different campaigns: the 2013A, 2014A and 2015A. The 2013A campaign observed  40 DECam fields (120 deg$^2$) every 2 hours (exposures of 173 s) during 4 nights in $u'$ band. In 2014A we observed 40 DECam fields (120 deg$^2$) every 2 hours (exposures of 160 s) during 5 nights in $g'$ band. The 2015A campaign consisted of 6 consecutive nights surveying 50 DECam fields (150 ${\rm deg}^2$) with a cadence of 1.6 hours (exposures of 87 s) in $g'$ band. The 2015A data are not as deep as in the 2014A campaign but survey a wider area and increase the number of visits per night from 5 to 6. These 6 nights were followed by three nonconsecutive half nights 2, 5 and 20 nights after the end of the main run. Some of the DECam pointings during the 2015A campaign were observed in $r'$ and $i'$ bands, but not more that once per night (and only in a few nights). The details of HiTS can be found in \cite{2016ApJ...832..155F}, along with a comparative table of its three campaigns. In this work we used data from the 2015A campaign and results from the 2014A campaign \citep{Pena.2018}. Since HiTS was not designed for asteroid observations, all asteroids observations were serendipitous. In 2014A the observations reached the ecliptic, while 2015A observations are at least $5^{\circ}$ away from the ecliptic. The ecliptic distribution of the observations in both campaigns is shown in Figure \ref{fig:ecliptic}.

\subsection{Data Processing} \label{sec:dataproc}
The data used for this work were processed and reduced in the same way as done for the 2014A campaign (see \citealt{2016ApJ...832..155F} and \citealt{Cabrera.2017}), meaning that we had astrometry and photometry of moving objects and a probability for each of them of being real or bogus obtained using deep learning.

An incongruity in the way reduced data from a few epochs for the 2015A campaign were stored produced some uncertainty in the actual observation time for detections in those epochs. To remedy this issue, we took advantage of a new reduction (using another pipeline, see section 3 of \citealt{Martinez.2018}) that had the correct observation times but without their probability of being real or bogus.
We took the old data (with possible inconsistencies in their observation time) and we matched their positions with those detections of the new reduction; so if more than fifty percent of detections were within 2 pixels from the old data set to the new one for a given exposure, we considered that the data from that exposure had the correct observation time stored and we kept using those old detections, allowing us to discriminate between them to keep those with high probability of being real. After this \emph{quality control}, we decided not to use data from 4 entire fields and 13 exposures from different fields that showed too large distances between detections from the new and old databases or have not enough detections to compare. Two of those rejected fields were among the closest to the ecliptic. Finally we ended up with 154,444 detections of moving objects with a probability of being real higher than 0.5. %from 47 DECam fields. 
This is less than half of detections from the 2014A campaign. Although for 2015A we had data of 7 more fields than in 2014A, the fewer data can be explained by the limiting magnitude of 0.5 to 1 magnitudes brighter (because of the smaller exposure time and bad weather) and because in 2015A the observations were at least $\sim6$ degrees away from the ecliptic while 2014A data reached it.

\subsection{Survey Efficiency}
As for the 2014A data, we used the estimated position of known asteroids to test the asteroid detection efficiency of the HiTS 2015A survey. By checking the number of times a known asteroid is detected as a variable source we got an accurate assessment of the maximum number of asteroids our linking algorithm can identify as a moving object.

\begin{figure}
   \centering
   \includegraphics[width=\hsize]{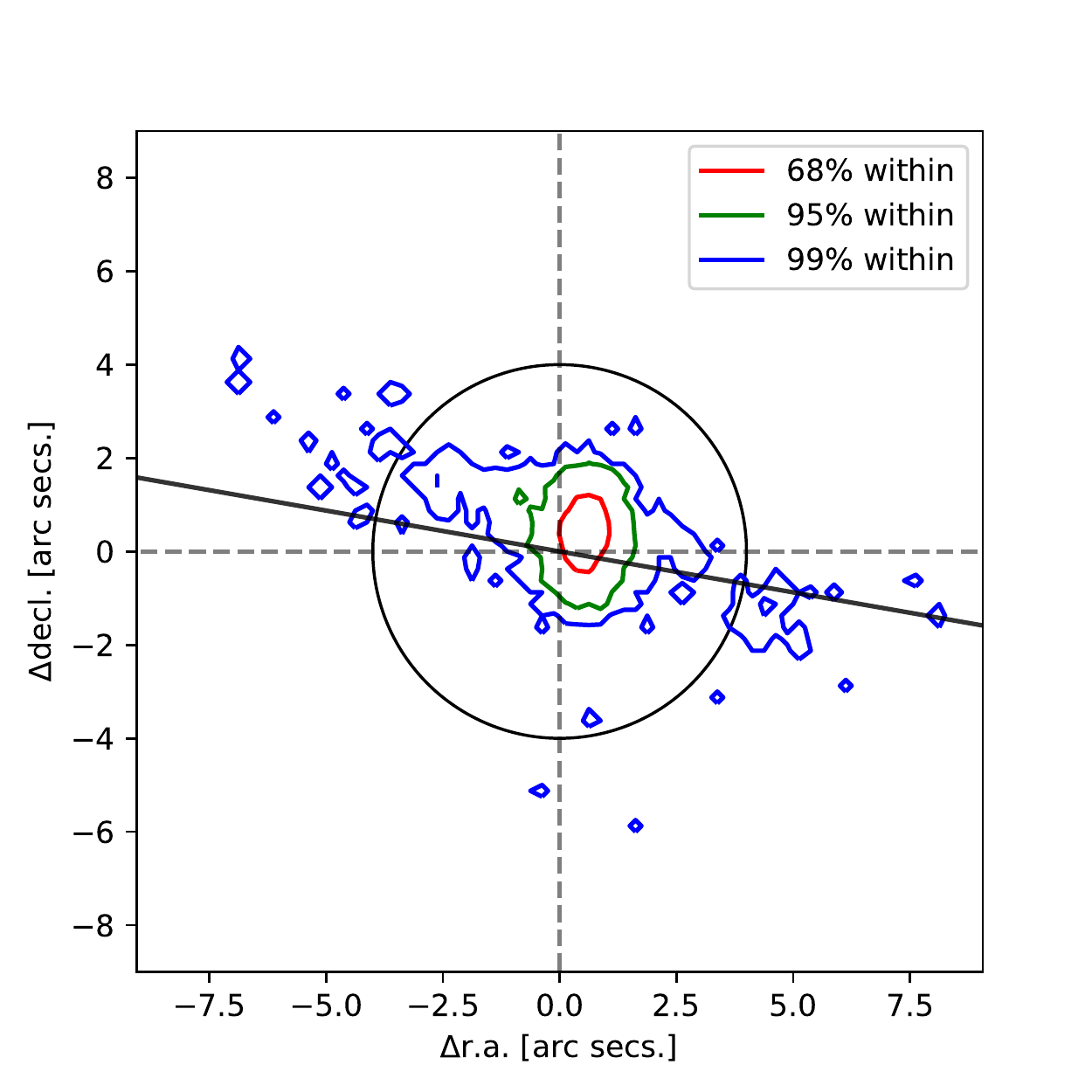}
   \caption{Contour plot for the difference between coordinates of JPL and HiTS data. Each contour surrounds a percentage of the matched data. The straight black line has the same slope as the ecliptic. The black circle has a $4\arcsec$ radius.}
   \label{fig:distcontour}%
\end{figure}

But first we had to decide how far a detection can be from the estimated position of a known body to consider it as \textit{``recognized.''} We had previously used information from the Minor Planet Center\footnote{Information provided via web page at \url{https://www.minorplanetcenter.net/cgi-bin/checkmp.cgi}} (MPC) to look for bodies within 1.25 degrees of any DECam pointing in HiTS 2015A, but the necessity of having updated coordinates of these bodies led us to use the Jet Propulsion Laboratory (JPL) web service\footnote{JPL Horizons: \url{https://ssd.jpl.nasa.gov/horizons.cgi}} (which gave us computational simplicity for big queries) to get the coordinates of those same asteroids. Using the coordinates of known bodies obtained this way, we computed the distance between JPL and HiTS detections, yielding the distribution shown in Figure \ref{fig:distcontour}. Is remarkable that using data from JPL we concentrated the difference to a much smaller range, allowing us to easily reduce our criteria for recognizing detections to a $4\arcsec$ distance (in comparison to the $7\arcsec$ in \citealt{Pena.2018}).

\begin{figure}
   \centering
   \includegraphics[width=\hsize]{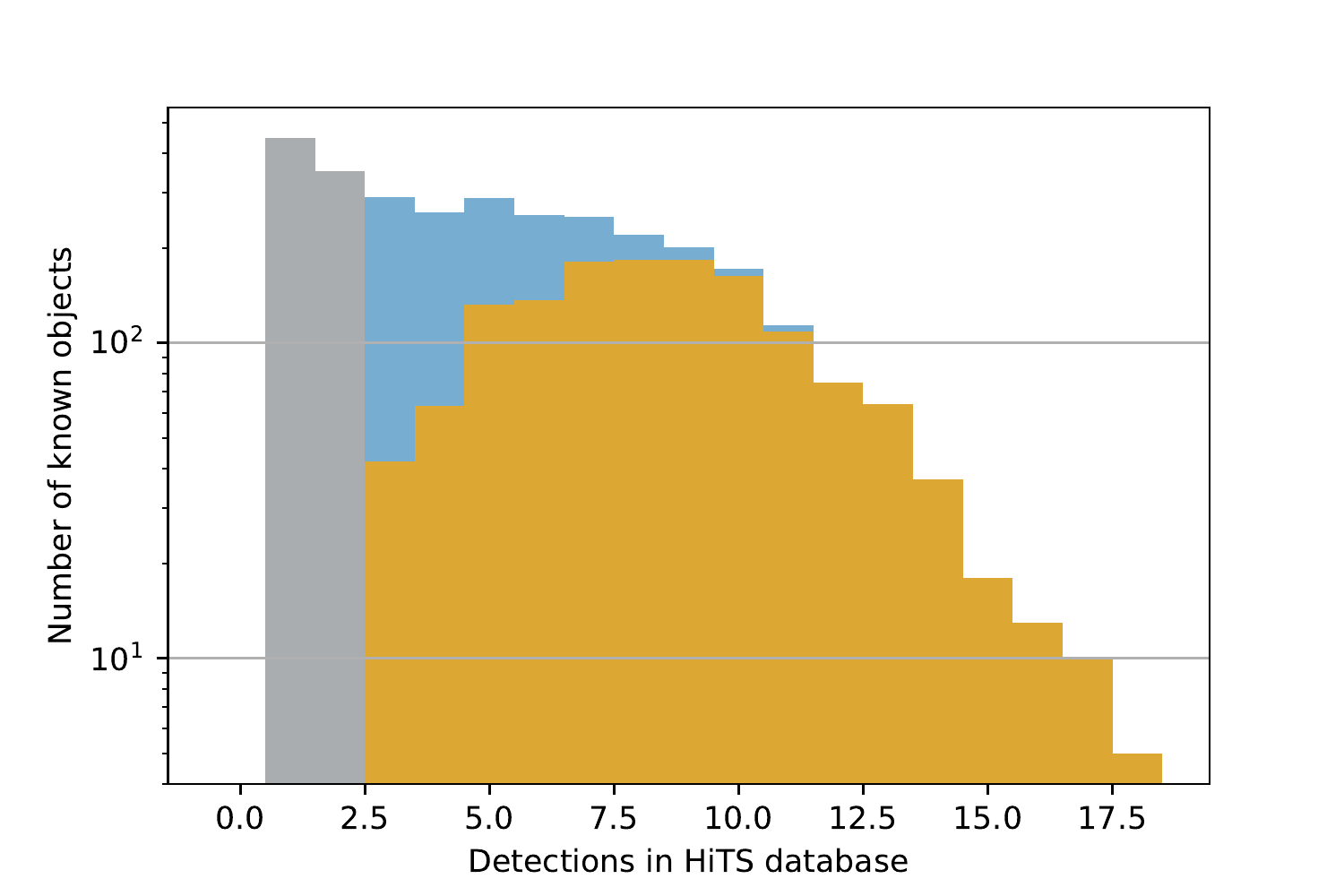}
   \caption{Histogram of number of times a known asteroid was imaged and analyzed as a variable candidate source by the HiTS survey. All known asteroids (from MPC) that were within 1.25 degrees of any DECam pointing in the HiTS survey during 2015 were considered in this analysis (using coordinates delivered by JPL). Asteroids with fewer than three detections are shown in gray, those with three or more are colored blue and those with at least three detections in a single night are shown in orange.}
   \label{fig:detectedhist}%
\end{figure}

Using the recognized detections, we could see how many of the known bodies we found. In Figure \ref{fig:detectedhist} we show the number of asteroids as a function of the number of detections. In gray we show those that are found only in one or two detections and in blue those found three or more times. But since our linking algorithm required at least three detections per night (Section \ref{sec:analysis}), we show in orange the subset of asteroids satisfying that condition.

\section{Analysis} \label{sec:analysis}
Since the detections with a probability of being real higher than 0.5 were more sparse than those in the 2014 campaign, linking detections for one night to another proved to be harder than in \cite{Pena.2018}. To solve this, we first found \textit{tracklets} (sets of at least three detections that assimilate a linear trajectory in \textit{one} night). To link different tracklets between nights we tried three different algorithms. Finally we got a collections of \textit{tracks} (composed by 1 or more linked tracklets).

\begin{figure}
   \centering
   \includegraphics[width=\hsize]{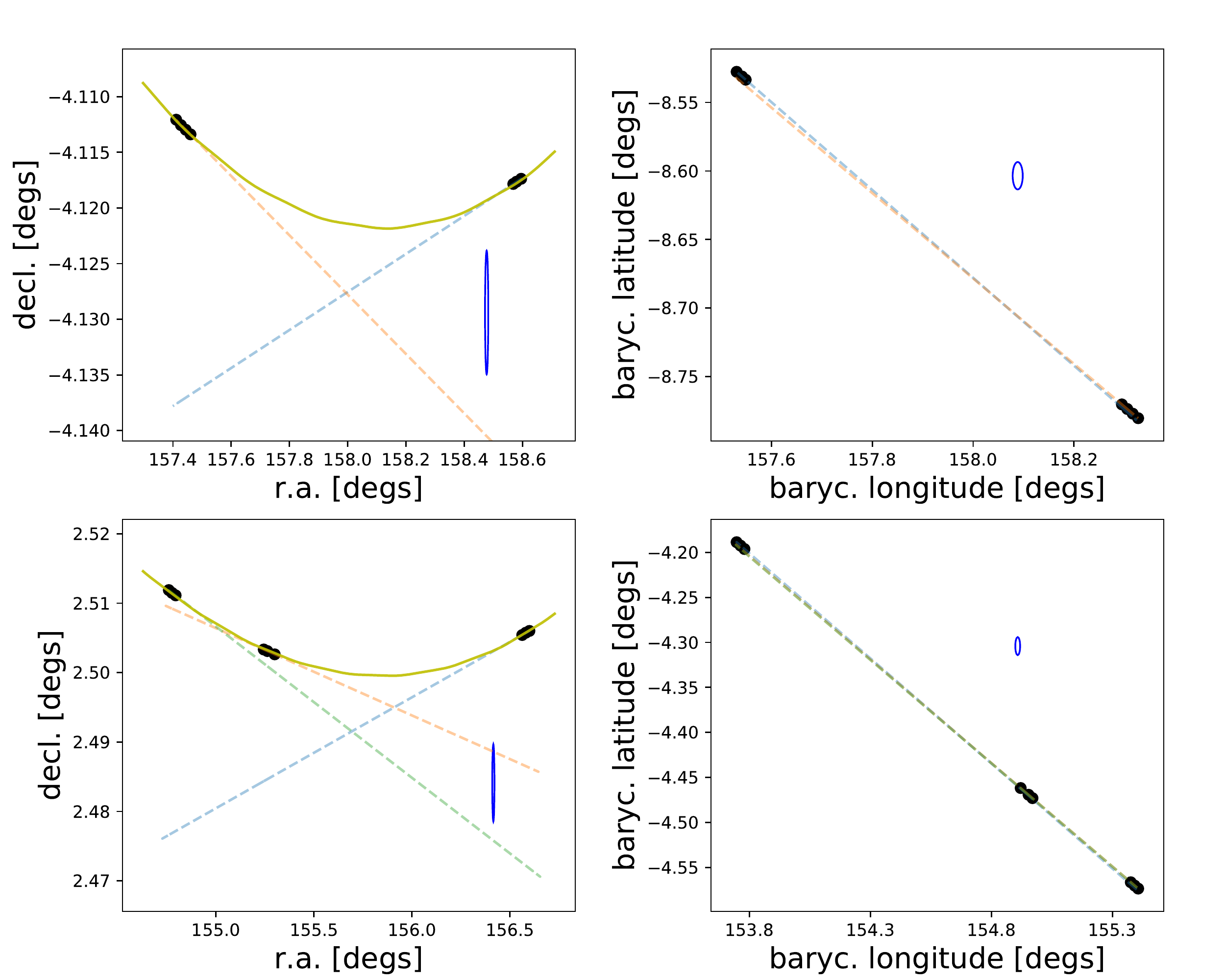}
   \caption{Known asteroids with difficult trajectories. The upper two panels show the asteroid 2009 CP21 and the lower two panels the asteroid 2017 SJ103. On the left, in equatorial coordinates,  the HiTS detections (black dots) and the trajectory as estimated by JPL Horizons (yellow line) are shown. On the right we show in black dots the barycentric coordinates as estimated assuming barycentric distances similar to the actual ones. Dashed lines show linear fittings for all tracklets and the blue circles show the maximum distance between estimated coordinates to join these tracklets into tracks (20$\arcsec$ for the first linking algorithm working in equatorial coordinates and 36$\arcsec$ for the last algorithm working in barycentric coordinates).}
   \label{fig:weird_asts}%
\end{figure}

In the first algorithm we tried, we took pairs of tracklets. If their 
%(linearly or quadratic if possible) 
estimated position in three different times (conveniently chosen for each pair to fall in the middle of them and near each tracklet) fell near each other, then we joined these tracklets.
The positions were estimated using quadratic fitting of the tracklets. How far apart the estimated positions could be depended on how far they were from the tracklets (until a maximum distance of $\sim20\arcsec$). But finally this algorithm failed to link several known asteroids (such as the ones shown in Figure \ref{fig:weird_asts}). Since we needed tracks to be in at least 2 nights to have good orbital parameter estimation (see section \ref{sec:orbfit}) we realized we needed a better algorithm.

The second algorithm we tried was HelioLink \citep{Holman.2018}, which takes the method shown in \cite{Bernstein.2000} but moves the coordinates origin to the barycenter (or to the Sun) and assumes a distance and a velocity of the asteroid with respect to this origin. But the high density of tracklets in the ($\theta_x,\theta_y$) space (equation 1 in \citealt{Holman.2018}) together with the clustering parameter (equation 11 in \citealt{Holman.2018}) that encloses two distances in a single parameter caused  many clusters to mix 
tracklets that did not belong together. %true asteroid's tracklets with alien ones. 

The third algorithm (and the one we finally used) uses a similar approach to that of HelioLink: assuming the same barycentric distance for all tracklets, clusters were made if the estimated barycentric ecliptic position coincided in two different times (the detailed algorithm is shown in Appendix \ref{appendix:1} and \ref{appendix:2}). Moving the coordinate reference to the solar system barycenter allowed us to cluster tracklets estimating their positions using linear fitting, as seen in Figure \ref{fig:weird_asts}, where curved trajectories as seen from Earth (left panels) are seen as straight lines (right panels). To cluster as many tracklets as possible it was necessary to try different barycentric distances, although most of them were obtained assuming a distance of 2.5 au (roughly the middle of the Main Belt). With this method we found 1770 asteroids detected in more than 1 night (which highly increased the orbital parameters accuracy, see section \ref{sec:orbfit}) while using the first method we only found $\lesssim$1100.

\begin{figure}
   \centering
   \includegraphics[width=\hsize]{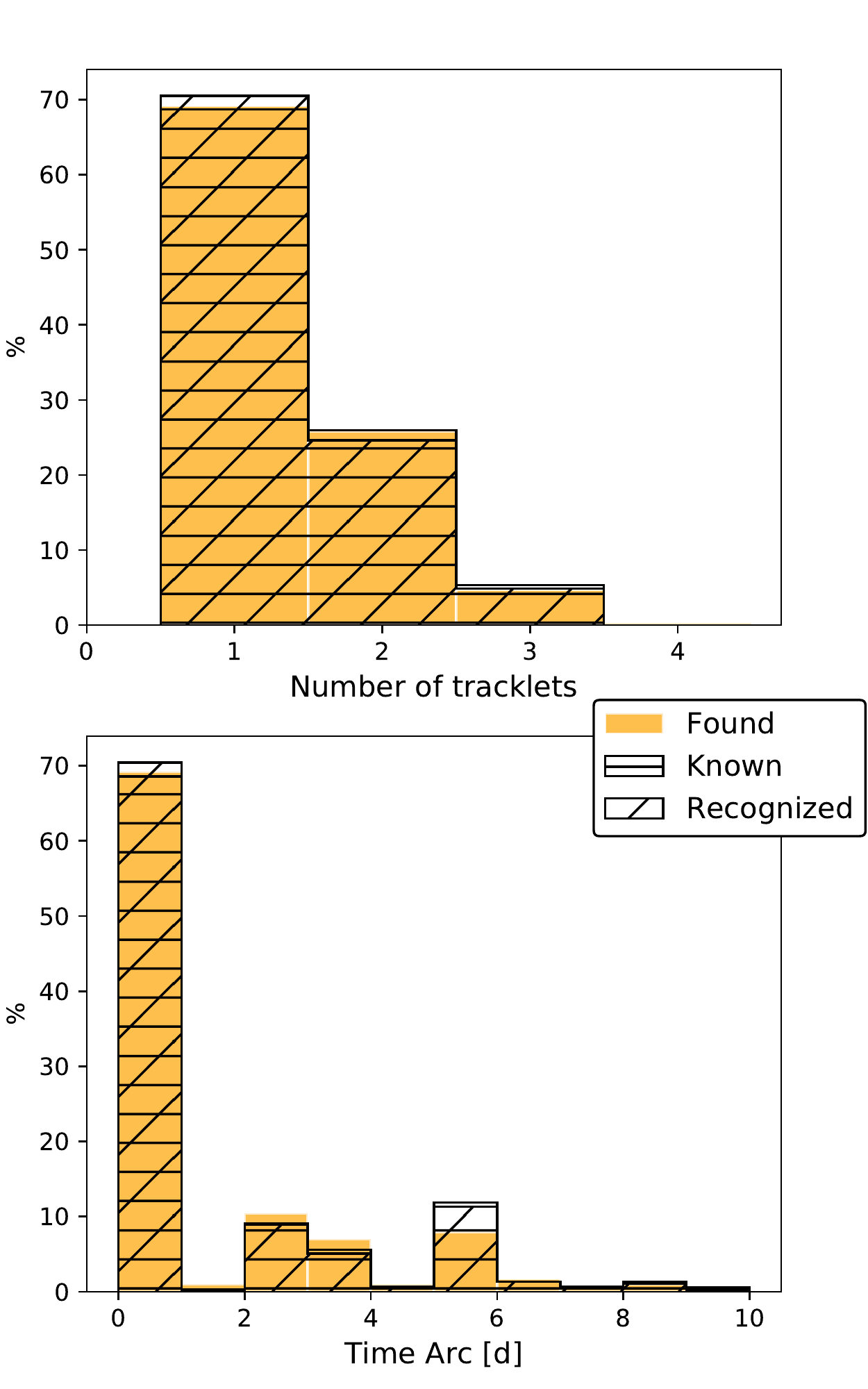}
   \caption{Upper panel: percentage of clusters grouped by their number of tracklets. Lower panel: percentage of clusters grouped by their time arc. In yellow all found clusters are shown, the known ones --if perfectly linked-- are shown with horizontal lines, and as actually found with diagonal lines. In total there are 7945 tracklets grouped in 5674 clusters (yellow), 1324 known bodies among those tracklets (horizontal lines) that were found in 1403 clusters (diagonal lines).}
   \label{fig:clusters}%
\end{figure}

To prove that the clustering works properly, we compared the total amount of clustered tracklets with the amount of tracklets that could be obtained from the known asteroids. In Figure \ref{fig:clusters} we show the number of tracklets per cluster (upper panel) and the time arc per cluster (lower panel) for all found clusters, for all known clusters (tracklets identified as known bodies) and for known clusters as they were actually found by our algorithm (the \textit{recognized} ones), in percentage in each case. Although $\sim70\%$ of tracklets were not linked with any other (clusters of 1 tracklet), the same happened with the known clusters. This, together with the fact that we could link almost all known clusters without contamination (without wrongly joined tracklets), we consider that our linking algorithm worked very well for our data. %(filtered data using Random Forest covering a time arc of mainly $\sim$5 days for most data and with a few observations of at most $\sim$10 days).

\section{Results} \label{sec:results}
We produced a total of 5740 tracks after the clustering process. We checked the efficiency of our analysis with the 1422 known objects (Figure \ref{fig:distcontour}) that our process could have linked (at least three recognized detections in one night, orange in Figure \ref{fig:detectedhist}). We found 1323 objects distributed in 1349 tracks. This means we failed to link a minority of related tracklets, yielding a 93\% detection efficiency. %(see Figure \ref{fig:maghist}).

\subsection{Orbital Fitting} \label{sec:orbfit}

\begin{figure}
   \centering
   \includegraphics[width=\hsize]{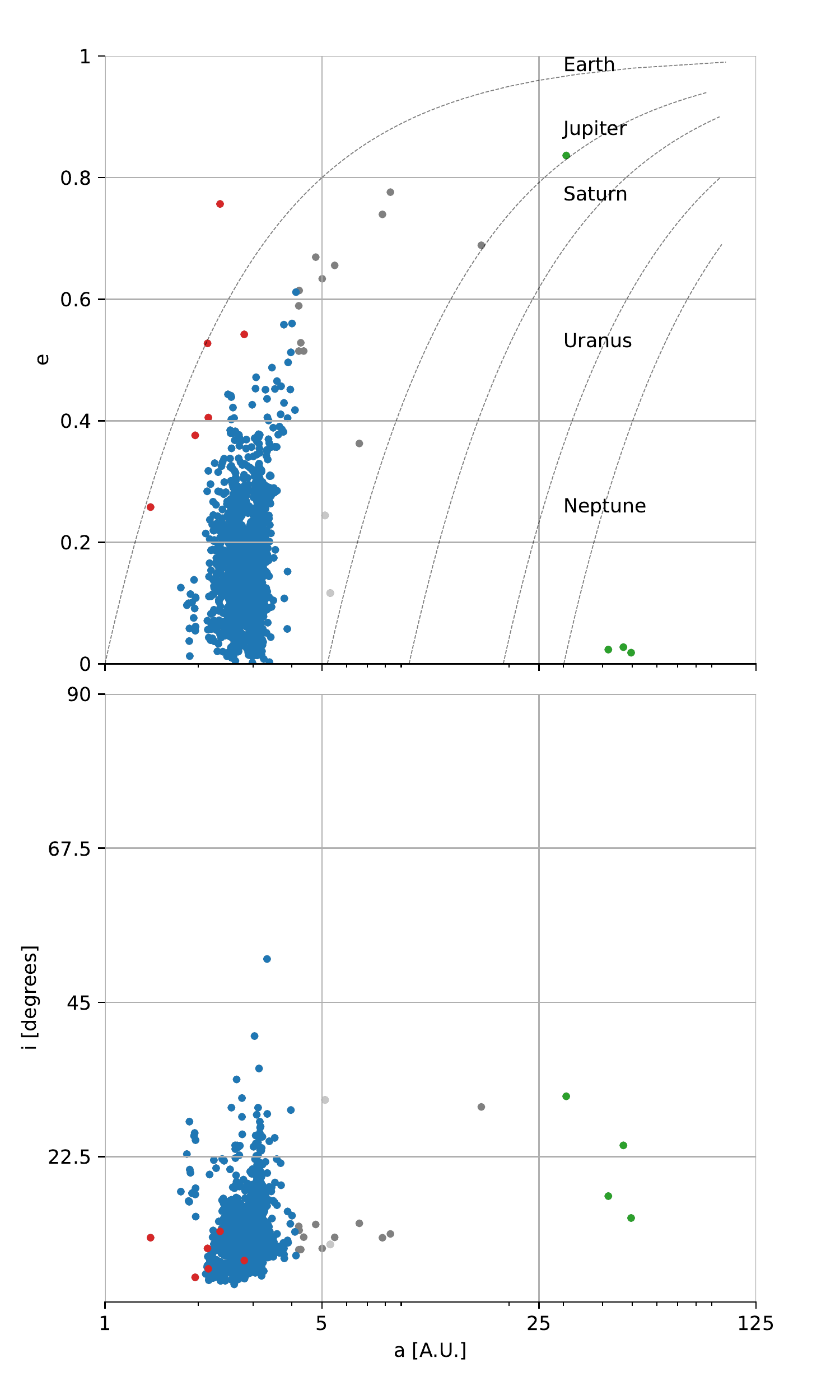}
   \caption{Orbital solution for all tracks detected in at least two nights that yield bound orbits and a maximum deviation of $2\arcsec$ from the model. There are 1762 objects in 2015 that fulfill this criterion. The lines show the solutions that share their pericenter distance with the outer planets. We show Near Earth Objects in red, Main Belt asteroids in blue, Trans-Neptunian objects in green and others in gray. No Centaurs were found.
   }
   \label{fig:orbits}
    \end{figure}

As in \cite{Pena.2018}, we applied a Keplerian orbit fit to each track. Due to the degeneracy in distance and velocity for tracks that span only a few hours we focused only on those that include observations in different nights. % ($>$0.5 days).
We further rejected all trajectories that yield unbound solutions or that deflect more than $2\arcsec$ from an observation, leaving 1762 bound trajectories or \textit{good tracks} from now on. We found 1738 Main Belt asteroids, 6 Near Earth Objects (NEOs) and 4 TNOs as defined in \citealt{Pena.2018} (See Figure \ref{fig:orbits}). 

\begin{figure}
   \centering
   \includegraphics[width=\hsize]{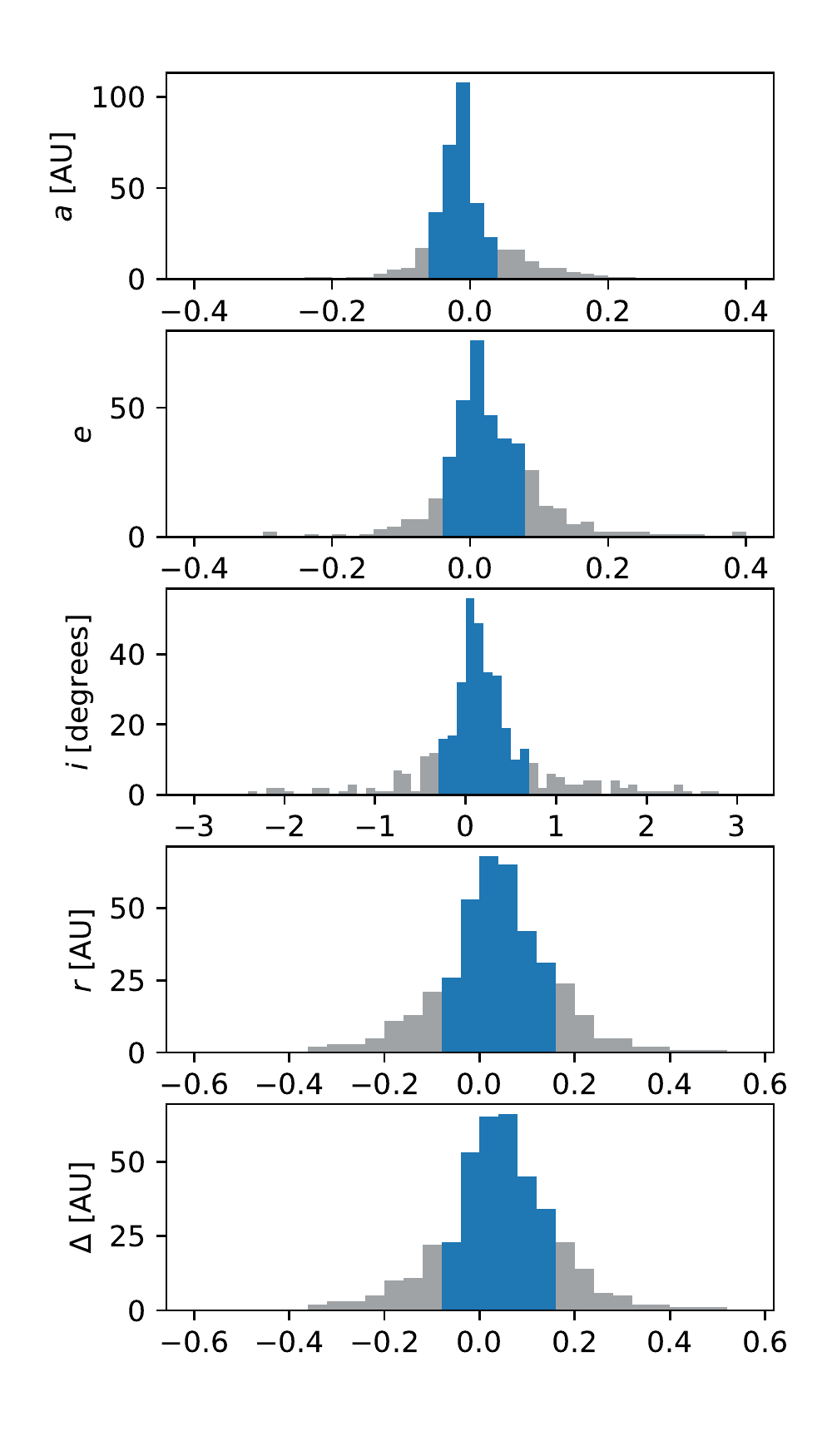}
   \caption{Errors for the estimated orbits of known asteroids in our sample detected in more than one night and with bound computed orbits with fitting errors of $\leq2\arcsec$. We highlight the $1\sigma$ confidence region (in blue). From top to bottom we show errors in semi-major axis $a$, eccentricity $e$, inclination $i$, barycenter distance $r$ and observer distance $\Delta$. The implied $1\sigma$ confidence region for our orbital solutions is: $\sigma_a = [-0.06, 0.04]~{\rm au}$, $\sigma_e = [-0.04, 0.08]$, $\sigma_i = [-0.3, 0.7]~{\rm degrees}$, $\sigma_r = [-0.08, 0.16]~{\rm degrees}$ and $\sigma_{\Delta} = [-0.08, 0.16]~{\rm degrees}$.
}
\label{fig:orberrors}%
\end{figure}

%\subsection{Orbital Parameters Errors}
We estimated our orbital parameter uncertainties using the detection of the 397 known objects with \textit{good tracks} recognized in our sample (see Figure \ref{fig:orberrors}). We report our $1-\sigma$ uncertainties as the interval that bounds 68\% of the errors around the mode (as in the normal distribution) to be $\sigma_a\sim 0.05 ~{\rm au}$ for the semi-major axis, $\sigma_{\rm e} \sim 0.06$ for the eccentricity, $\sigma_i\sim 0.5~ {\rm deg}$ for the inclination, $\sigma_{\rm r} \sim 0.12~ {\rm au}$ for the body-barycenter distance and $\sigma_{\Delta}\sim 0.12~ {\rm au}$ for the body-observer distance.
%To clarify our criteria, if we constrain further the Keplerian fit residuals, the improvement in errors are marginal and if we allow tracks with only one night, the tails in the errors distributions are highly increased.

\begin{figure}
   \centering
   \includegraphics[width=\hsize]{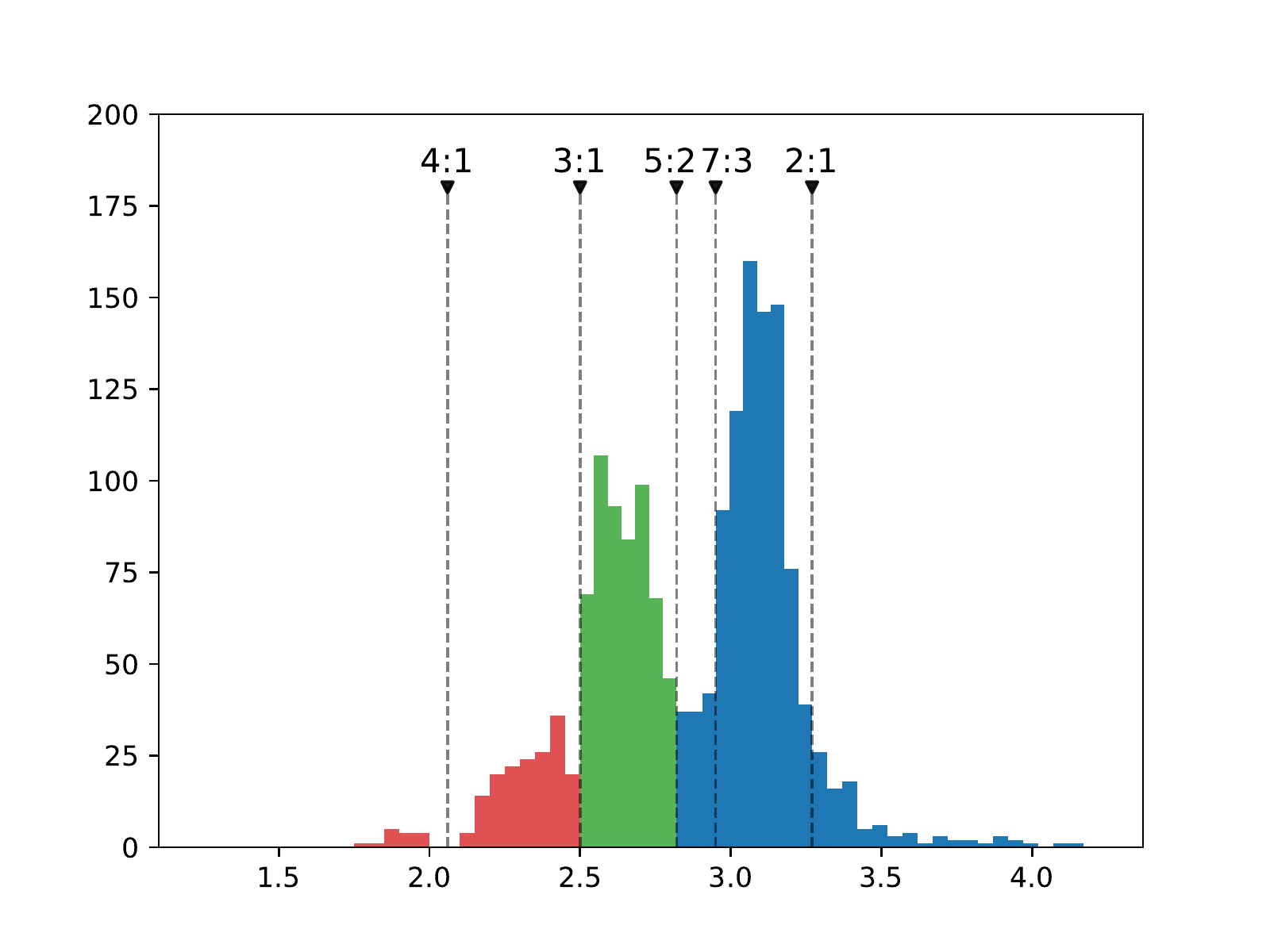}
   \caption{Main Belt distribution as a function of semi-major axis for the 1,738 bodies with \textit{good} orbital solutions (see Figures \ref{fig:orberrors} and \ref{fig:orbits}). Important Kirkwood gaps for resonances 4:1, 3:1, 5:2, 7:3 and 2:1 with Jupiter are plotted in dashed lines, %\sout{(semi-major axis of 2.06, 2.5, 2.82, 2.95 and 3.27 au respectively)}
   defining the Inner Belt (red), the Intermediate Belt (green) and the Outer Belt (blue), each one with 181, 566 and 991 bodies, respectively.}
% this binning is ad-hoc for each region within gaps... however bin difference is below 0.03 au
   \label{fig:MB}
    \end{figure}

In Figure \ref{fig:MB} we can see the distribution of the 1738 bodies identified as Main Belt objects divided in three groups: Inner Belt (from 1.3 to 2.5 au); Middle Belt (from 2.5 to 2.82 au) and Outer Belt (from 2.82 to 4.2 au). The limits between each group are in the most notorious Kirkwood gaps in Figure \ref{fig:MB} and have been used to differentiate the MB in different works (such as \citealt{Parker.2008, Masiero.2011} and \citealt{DeMeo.2013, DeMeo.2014}). In each of these populations we found 181, 566 and 991 bodies respectively.

\subsection{Magnitude Distribution}\label{sec:mag}

\begin{figure}
   \centering
   \includegraphics[width=\hsize]{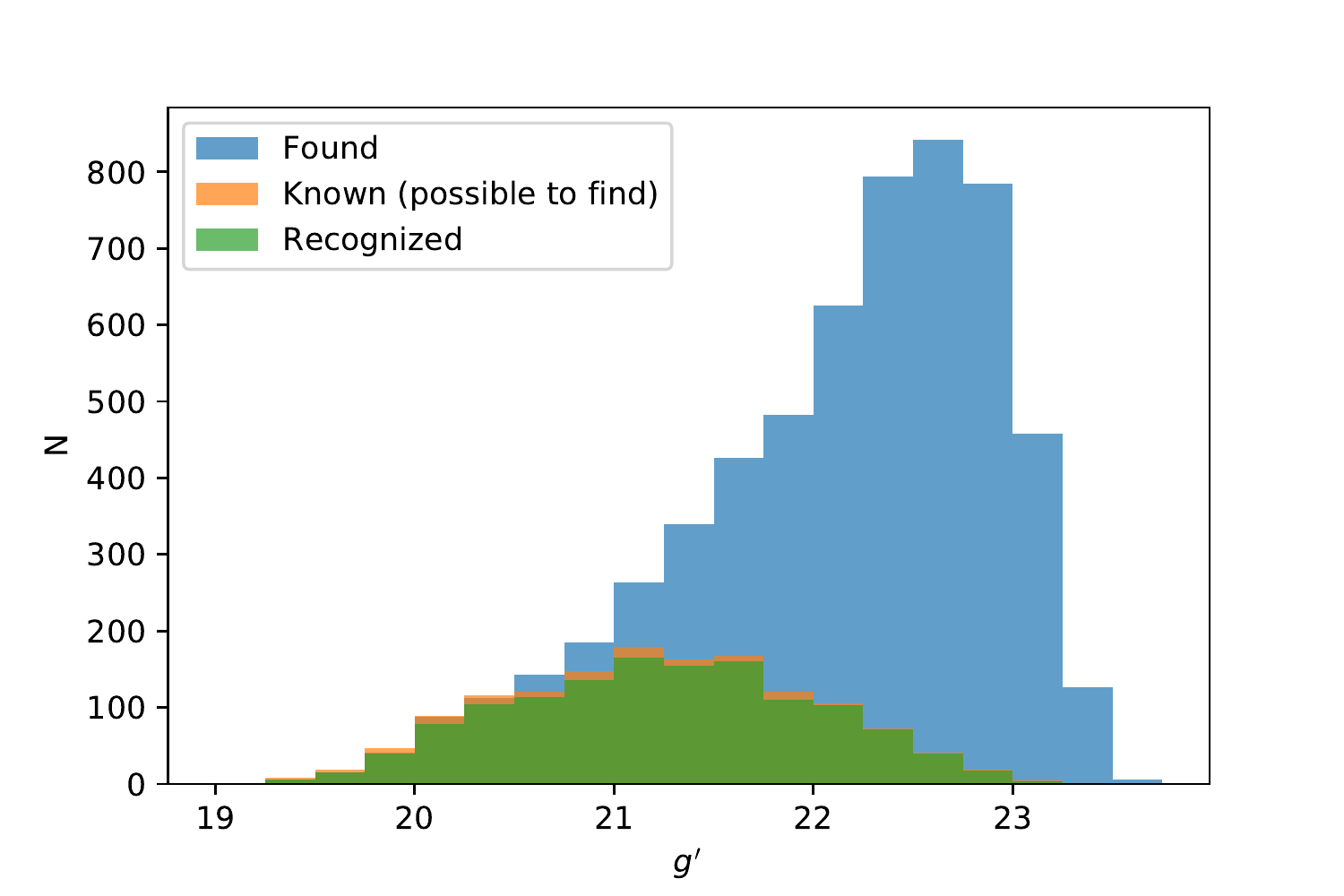}
   \caption{Histograms of the number of asteroids per magnitude (g' band). In orange, known asteroids (from JPL) that our linking process can find (wich haveat least three detections in any night). In blue, tracks found in the HiTS data. In green, those tracks that were recognized as known objects (at least three detections within $4\arcsec$ of a JPL detection). The drop in the blue histogram appears consistent with Figure~7 in \cite{2016ApJ...832..155F} once the effect of image subtraction is taken into account, which results in a loss of $\sim$0.4 mag. }
   \label{fig:maghist}
\end{figure}

%In Figure \ref{fig:maghist} we show distribution of t
For each track in our survey we computed a mean magnitude $g'$. In Figure \ref{fig:maghist} we show the distribution for all tracks in blue, for those that were recognized as known bodies in green and for all known bodies in orange (regardless of whether they were linked or not). %(for bodies found in more than one track, we show the mean of the joined magnitudes of those tracks). 
%In orange we show the distribution of magnitudes for all the known bodies that we should be able to find (i.e., they have at least three recognized detections in one single night). For those that were actually recognized, we use the mean magnitude of the track (from the green histogram) so the green and orange histograms are consistent, avoiding possible changes of the mean magnitude if the tracks do not include all the detections of the known bodies. Our limiting magnitude is under the magnitude 23, which is consistent with the limiting magnitude found in \cite{2016ApJ...832..155F} after taking into account a loss of $\sim$0.4 magnitudes because of the image subtraction process. %getting a proportion of faint bodies much greater than the known ones. %\jpz{\emph{Tal vez quitar:} In Figure \ref{fig:eff} we see the fraction of recognized object per $g'$ magnitude bin (bin fraction between green and orange histograms of Figure \ref{fig:maghist}). We were able to find the 93\% of the known bodies' tracks with very similar lengths to the actual ones (Figure \ref{fig:clusters}).}
%\begin{figure}
%   \centering
%   \includegraphics[width=\hsize]{f4.pdf}
%   \caption{\cfg{Perhaps have at most 10 bins} Efficiency of recognized objects over the known objects per magnitude bin (see green and orange histograms in Figure \ref{fig:maghist}). Errors are propagated using Poisson errors for each measurement. The dotted red line represents the total efficiency (total number of recognized tracks over the total number of known objects), equal to 0.93.\jpz{Cambiar de Poisson a Bernoulli.}}
%\label{fig:eff}
%\end{figure}

\begin{figure}
   \centering
   \includegraphics[width=\hsize]{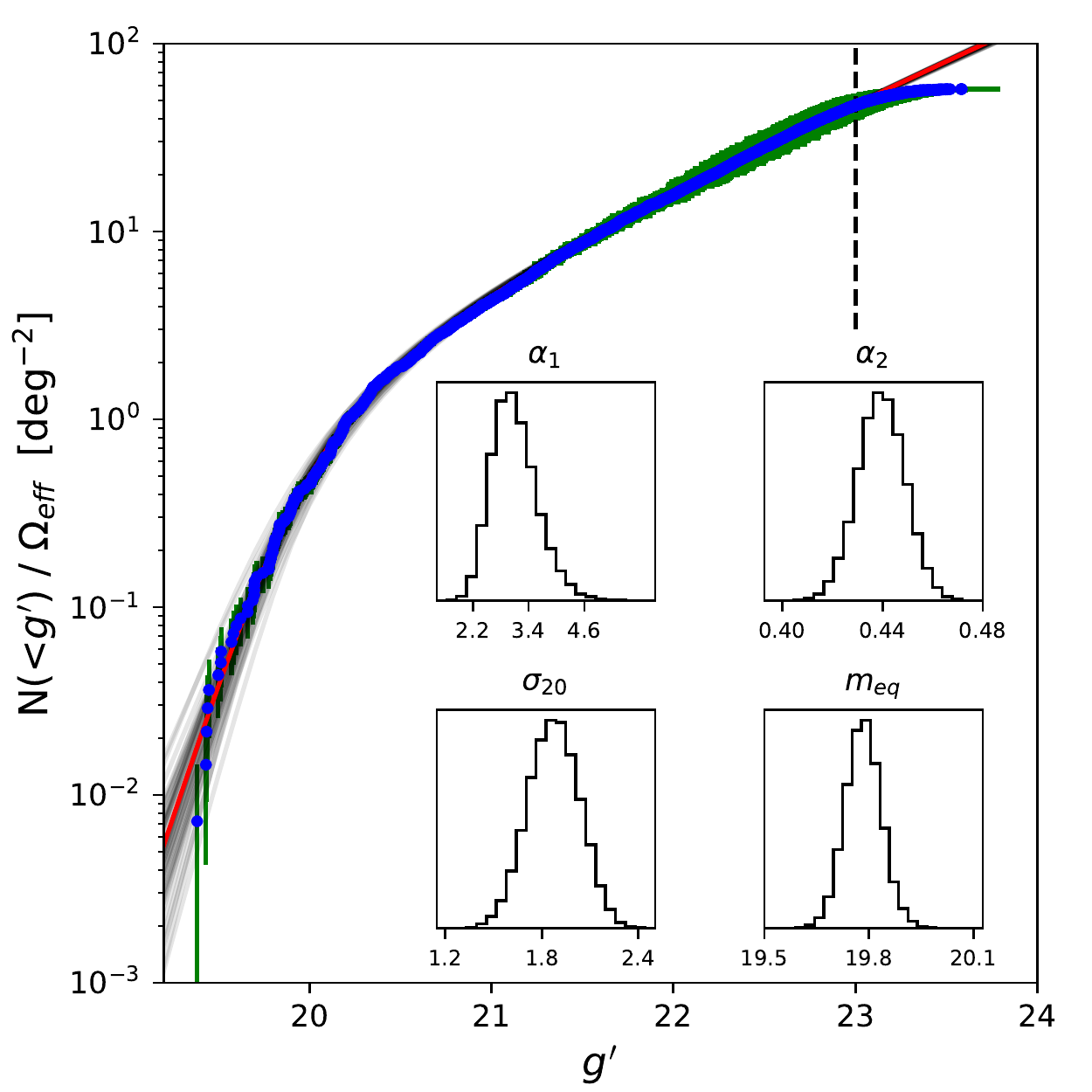}
   \caption{Cumulative distribution of magnitude $g'$ for the 5703 tracks with MB-like velocities (see section \ref{sec:mag}). A DPL (equation \ref{eq:dpl}) was fitted using MCMC to the 5119 bodies brighter than 23 $g'$ (marked with a dashed line). The small panels show the DPL parameters distribution. Using the mode with a $\pm$34\% confidence interval, we got $\alpha_1 = 2.94^{+0.48}_{-0.48}$, $\alpha_2 = 0.44^{+0.01}_{-0.01}$, $\sigma_{20} = 1.86^{+0.22}_{-0.15}$ and $m_{eq} = 19.78^{+0.06}_{-0.06}$. The red line shows the DPL given by the median values and the gray lines show 50 random models from the MCMC procedure. }
\label{fig:cumg}
\end{figure}

In Figure \ref{fig:cumg} we show the luminosity function (LF) %cumulative distribution (CD) of
in the apparent $g'$ magnitudes for almost all tracks. We left out 37 that showed non-MB orbits considering a criterion similar to that in \cite{Yoshida.2003} and \cite{Yoshida.2007} (namely, only tracks with MB-like ecliptic velocities of
%these by with MB-like velocities  this is velocity along the ecliptic under 
$<-0.15^{\circ}\mathrm{day^{-1}}$ were included\footnote{Obtained by linearly fitting geocentric ecliptic coordinates using \texttt{Python}'s (\url{https://www.python.org/}) \texttt{astropy} package (\url{https://www.astropy.org/}).}). %, for all tracks and using the known ones as references).
%This criterion is similar to the one used by%makes sense when compared with MB ecliptic velocities from
%\cite{Yoshida.2003} and \cite{Yoshida.2007}., where all populations appear separated when circular orbits are assumed, but when checking against known bodies, 4 of the 5 known NEOs (less than 0.5\% of all known bodies) appear well inside the MB velocity range. In consequence, 
We expect some contamination among those 5703 track from bodies outside the MB (specially from NEOs), but we do not expect it to be larger than 1\%.

The LF exhibits a very clear break that was fit with the harmonic mean of two power laws or double power law (DPL, see equation \ref{eq:dpl}) %to the magnitude surface density,
using the same method shown in \cite{Bernstein.2004}, \cite{Fuentes.2008} and \cite{Fuentes.2009}, based on the likelihood function derived by \cite{Schechter.1976}. %(equation \ref{eq:likelihood}, where $\sigma$ is the magnitude distribution, in this case from equation \ref{eq:dpl}, $\Omega$ is the surveyed area and $\eta$ is the magnitude efficiency).
\begin{equation}
\begin{aligned}
    \sigma(m) &= (1+c)\sigma_{20} \left[ 10^{-\alpha_1(m-20)} + c10^{-\alpha_2(m-20)} \right]^{-1} \\
    c &= 10^{(\alpha_2-\alpha_1)(m_{eq}-20)}
    \label{eq:dpl}
\end{aligned}
\end{equation}

%\begin{equation}
%    L(\sigma) \sim e^{-\Omega \int \eta(m) \sigma(m) dm} \prod_i \eta(m_i) \sigma(m_i)
%    \label{eq:likelihood}
%\end{equation}

We constrained the LF parameters in equation \ref{eq:dpl} using Python's package \texttt{emcee}\footnote{\url{https://emcee.readthedocs.io/}} \citep{Foreman-Mackey.2013} which applies an affine invariant Markov Chain Monte Carlo (MCMC) ensemble sampler \citep{Goodman.2010} that returns an approximation of the probability distribution as a function of the models' parameters. We considered a total survey area $\Omega=138 \deg^2$ and detection efficiency $\eta(m)$ as in \cite{2016ApJ...832..155F} (equation \ref{eq:eta}, with \text{erf} the error function\footnote{$\text{erf}(x)=\frac{2}{\pi}\int_0^x e^{-t^2}dt$}).
%but with one set of parameters for the entire survey, namely 
Taking into account %a drop of .5 magnitudes because
image subtraction and multiple detections yields parameters $m_{50}=23$  and $\Delta m_{50} = 1.1$ for the detection efficiency function of asteroids. 
\begin{equation}
    \eta(m) = \frac{1}{2} \left[ 1 + \text{erf} \left( -\frac{m-m_{50}}{\Delta m_{50}} \right)  \right]
    \label{eq:eta}
\end{equation}
We considered $\eta(m)$  and detections up to the limiting magnitude of our survey ($g'\sim$ 23), limiting our sample to 5119 objects. %brighter than our limiting magnitude ($g\sim23$)% for the parameters to converge.
Each of the 200 walkers used for this algorithm started at a random position near the parameter value obtained using the common $\chi^2$ minimization method. %\jpz{Esto por un comentario de otro autor.}
Using the 
%#$16^{th}$, $50^{th}$ and $84^{th}$ percentiles
mode with a $\pm 34\%$ confidence interval, we finally got $\alpha_1 = 2.94^{+0.48}_{-0.48}$, $\alpha_2 = 0.44^{+0.01}_{-0.01}$, $\sigma_{20} = 1.86^{+0.22}_{-0.15}$ and $m_{eq} = 19.78^{+0.06}_{-0.06}$.

This LF is very similar to that found by \cite{Gladman.2009}, with a break at $R\sim19$, consistent with our $m_{eq}\sim19.78$ (our mean color $g'-r'\sim$ 0.77, see Table \ref{tab:colors}). They got flatter slopes, especially at the bright end: $\alpha=0.61$ compared to our much steeper $\alpha_1\sim3.1$; while at the faint end they got $\alpha=0.27$ against our $\alpha_2=0.44$. The expected number of bodies is also lower for our survey: \cite{Gladman.2009} found $\sim90$ bodies per square degree brighter than $R\sim22$, while we only found $\sim30$ bodies brighter than $g'\sim22.7$. This is accounted by the fact that they pointed directly at the ecliptic %(being able to see bodies of any inclination),
while in this work the area closest to the ecliptic is at $\sim8^{\circ}$ with the bulk of our data is at $\sim15^{\circ}$. This is consistent with the results by \cite{Ryan.2009}, who showed that the number of detected asteroids decreases with ecliptic latitude by 50\% and 20\% at $10^{\circ}$ and $15^{\circ}$ with respect to $0^{\circ}$.

\subsection{Absolute Magnitude Distribution}\label{sec:H}
We computed absolute magnitudes $H$ for all \textit{good tracks}. In equation \ref{eq:H} $r$ is the body-barycenter distance, $\Delta$ is the body-observer distance, $\alpha$ is the phase (Sun-body-observer angle), and $\phi(\alpha)$ is the phase function %, which may take several different forms
(see \citealt{Waszczak.2015} for several definitions of $\phi$). We used the $(H,G)$ model for $\phi$ %given by equation \ref{eq:phi_alpha} 
\citep{Bowell.1989}, using the typical value of $G=0.15$ (as in the ephemeris data delivered by MPC and JPL). Since we had data mainly in $g'$ and some in $r'$, we computed the absolute magnitudes for each filter, $H_{g'}$ and $H_{r'}$. We report the average absolute magnitude for a track.

\begin{equation}
    H = V - 5\log_{10} (r\Delta) +  2.5\log_{10}[\phi(\alpha)]
    \label{eq:H}
\end{equation}
%\begin{equation}
%\begin{aligned}
%\phi(\alpha) &\equiv (1-G)\phi_1 + G\phi_2 \\
%\phi_1(\alpha) &\equiv \exp (-3.33 \tan^{0.63}[\alpha/2]) \\
%\phi_2(\alpha) &\equiv \exp (-1.87 \tan^{1.22}[\alpha/2])
%\label{eq:phi_alpha}
%\end{aligned}
%\end{equation}

\begin{figure}
   \centering
   \includegraphics[width=\hsize]{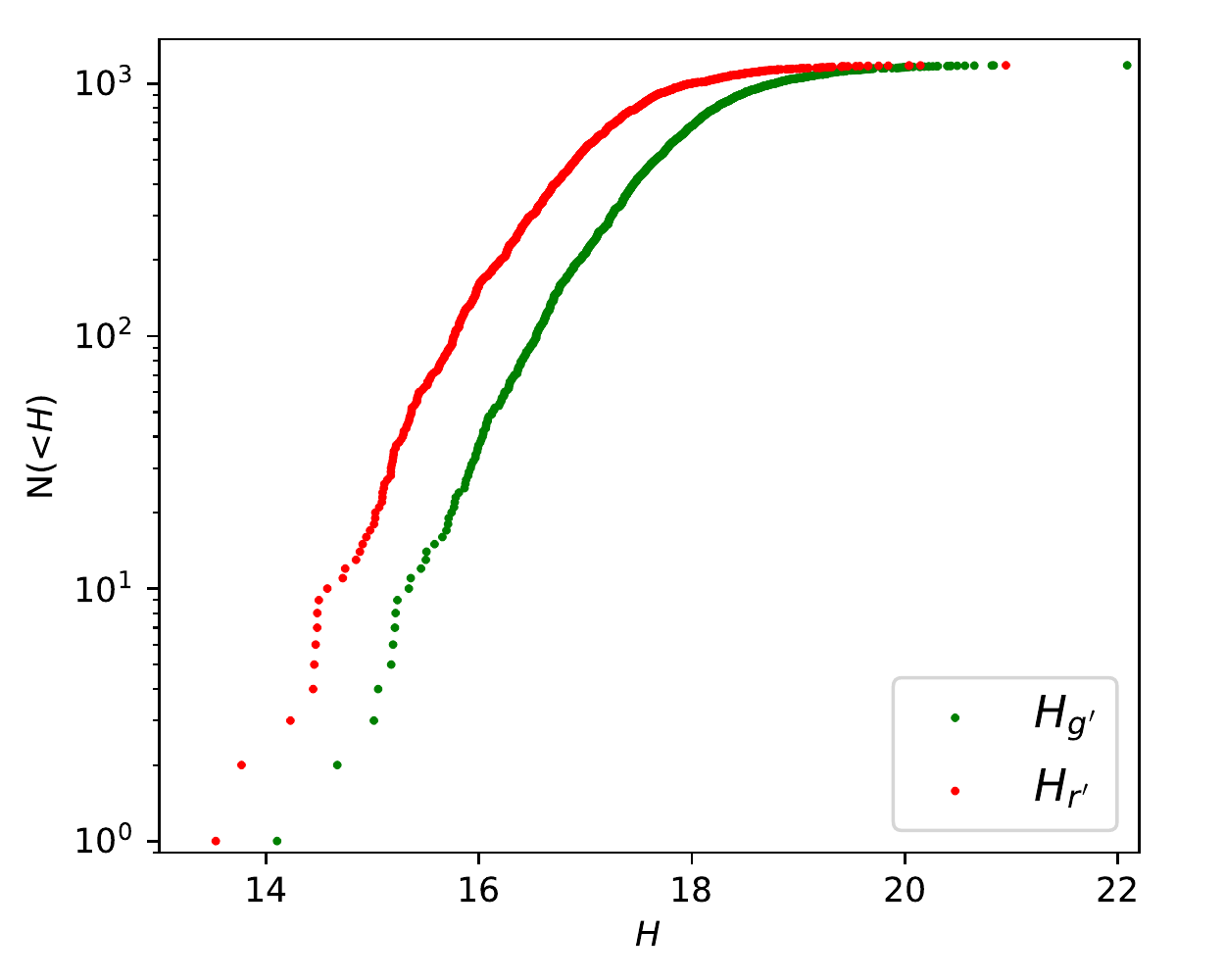}
   \caption{Cumulative distribution of $H_{g'}$ and $H_{r'}$ for the 1182 MB bodies with \textit{good} orbital solution and measurements in $g'$ and $r'$ bands. }
\label{fig:Hcolor}%
\end{figure}

Since $H$ can be related with the body's size by the equation $D=10^{-H/5}1329/\sqrt{p_V}$, where $D$ is the diameter in km and $p_V$ the geometric albedo, we sought for a possible color--size relation in the MB using $H$ as a proxy for the size (assuming a common $p_V$ for all bodies). In Figure \ref{fig:Hcolor} we show the cumulative \textit{size} distribution (CSD) for the 1182 MB bodies measured in $g'$ and $r'$ bands. Both $H_{g'}$ and $H_{r'}$ distributions are well represented by a single power law (SPL) with the same slope ($\sim$ 0.88), showing no evidence for any color--size relationship in our data. This result was unchanged even when we considered each MB zone separately (Inner, Middle and Outer).

\begin{figure*}
\gridline{\fig{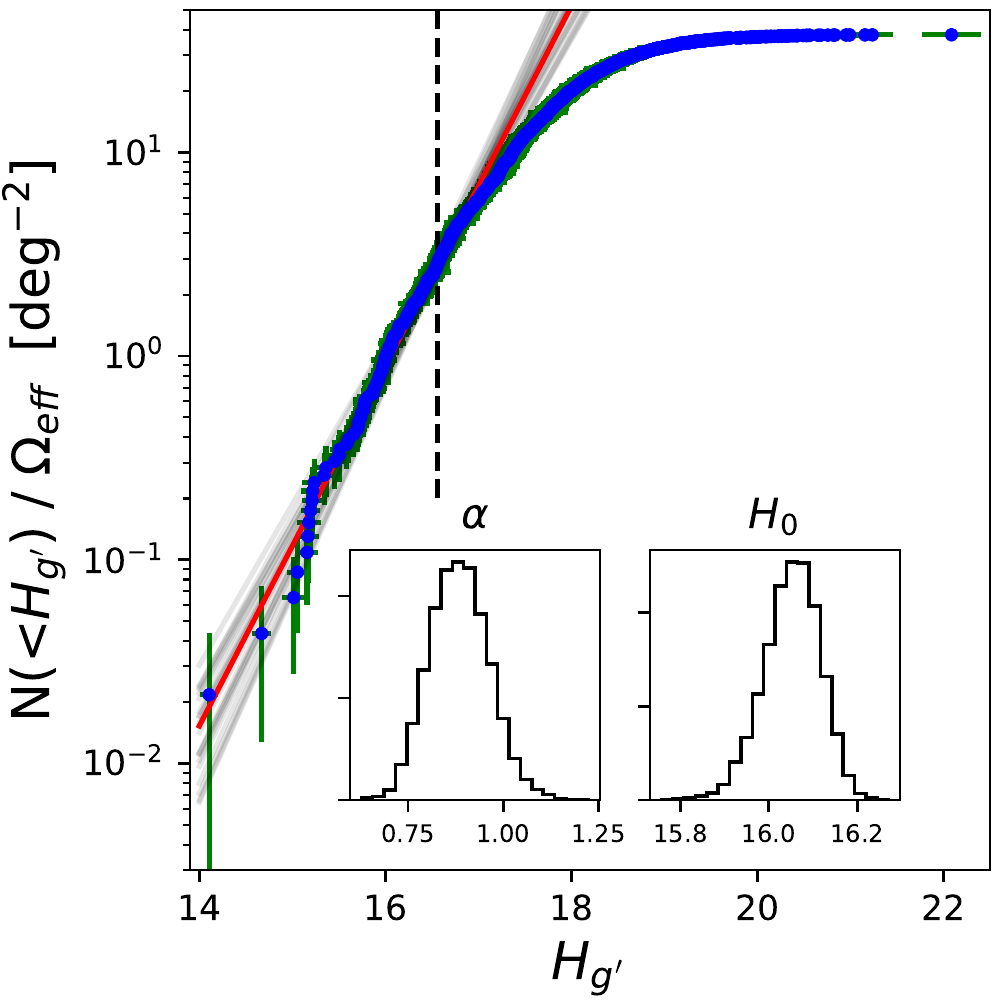}{0.25\textwidth}{(a)}
          \fig{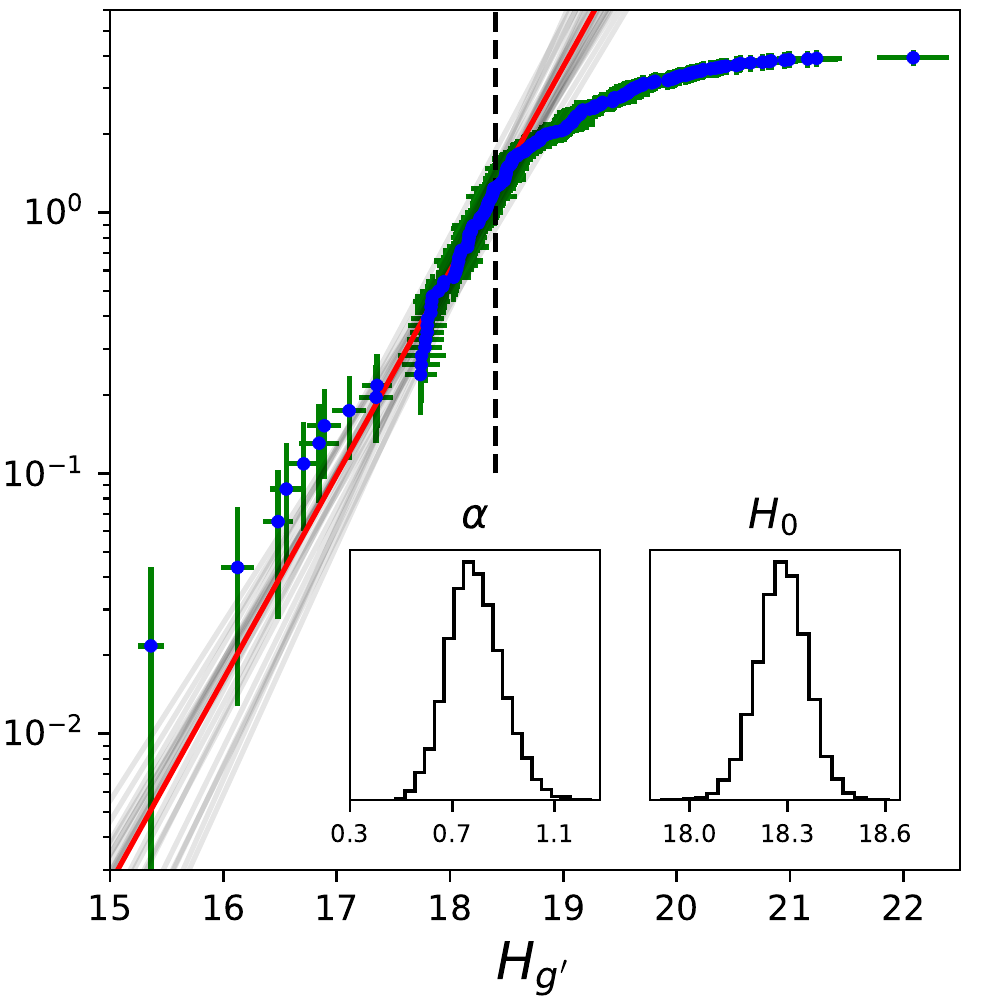}{0.25\textwidth}{(b)}
          \fig{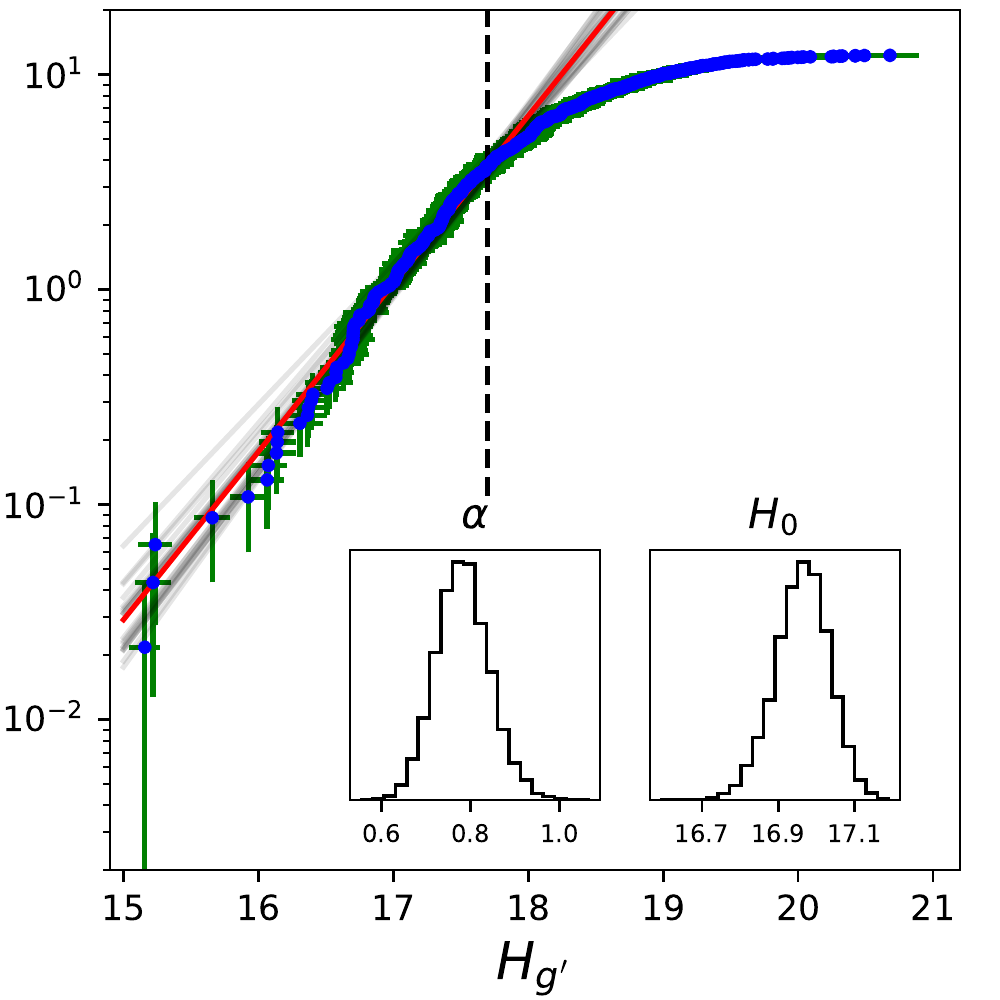}{0.25\textwidth}{(c)}
          \fig{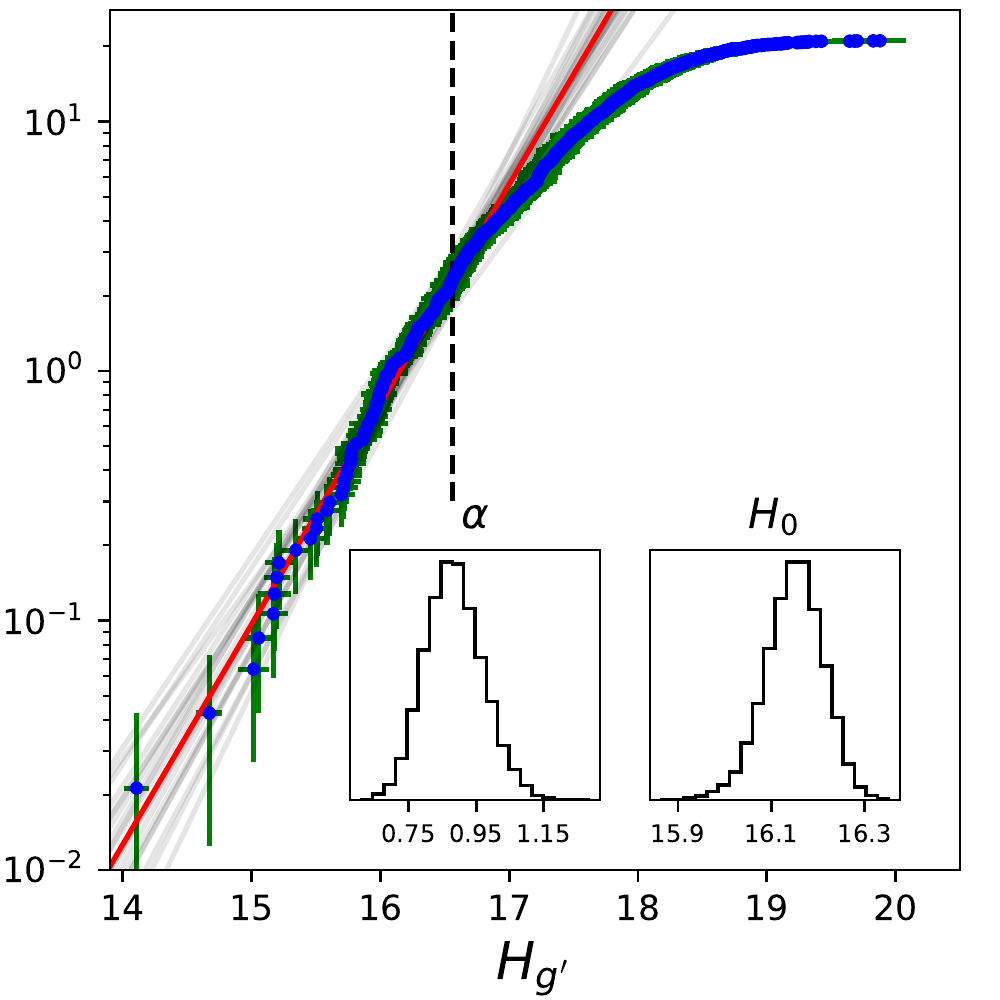}{0.25\textwidth}{(d)}
          }
\caption{Cumulative $H$ distribution of bodies with good orbital solutions for (a) all MB bodies; (b) inner MB bodies; (c) intermediate MB bodies; and (d) outer MB bodies. A single power--law distribution was fit ($\sigma(H_{g'}) = \alpha \ln(10) 10^{\alpha(H_{g'}-H_0)}$), with the two smaller panels showing the parameter distribution. In red we show the model given by the median values and in gray models given by 20 random values from the parameter distributions. %\cfg{why do we do this 20 random objects thing?}.
Dashed lines mark the limiting magnitude (only brighter bodies are considered for the fit). A summary of the fitting results is in Table \ref{tab:H}. 
\label{fig:Hdistr}}
\end{figure*}

%In Figure \ref{fig:Hdistr} is shown the cumulative distribution of our \textit{good} tracks assuming a constant discovery efficiency $\eta{H}$. 
In the following analysis we only considered $H_{g'}$ since all tracks were observed in $g'$.
Following the process described in section \ref{sec:mag}, we fit an SPL distribution $\sigma = \alpha\ln{(10)}10^{\alpha(mag-H_0)}$ using only bodies brighter than %a limiting magnitude given by 
our limiting magnitude ($g'\sim$ 23). %combined with the tracks elongations (Sun-observer-body angle) and their distances (see equation \ref{eq:H}) and assuming a constant discovery efficiency until that $H$ magnitude. 
The SDs are shown in Figure \ref{fig:Hdistr} and the most likely parameters %results of the fitting procedure 
are summarized in Table \ref{tab:H}. The MB as a whole or by subregion exhibits slopes much steeper than measured by other studies, only comparable to the ones measured by \cite{Parker.2008} for brighter objects. 

Since DECam has similar filters to those in SDSS  \citep{Schlafly.2018}, we used Lupton (2005)\footnote{\url{http://classic.sdss.org/dr7/algorithms/sdssUBVRITransform.html}} to transform $H_{g'}$ to $H$ ($g'$ to $V$). Lupton found that $V \sim g' - 0.58(g' - r')$ (similar to the results found by \citealt{Fukugita.1996} and \citealt{Krisciunas.1998}). Assuming all bodies have the same color and albedo, $g'-r'\sim$ 0.76 (see section \ref{sec:color}), and $p_V=0.1$ (roughly the mean albedo from \citealt{Polishook.2012}), sizes range between $1<D<10$ km for objects between $14<H_{g'}<18$.

\begin{deluxetable}{cCCCC}%[b!]
\tablecaption{Main Belt SPL fitting\label{tab:H}}
\tablecolumns{5}
%\tablenum{1}
\tablewidth{0pt}
\tablehead{
\colhead{Population\tablenotemark{a}} &
\colhead{$H_{lim}$\tablenotemark{b}} &
\colhead{N\tablenotemark{c}} &
\colhead{$\alpha$\tablenotemark{d}} &
\colhead{$H_0$\tablenotemark{d}} %\\
}
\startdata
Main Belt & 16.56 & 129 & 0.88^{+0.09}_{-0.08} & 16.04^{+0.09}_{-0.05} \\ \hline
Inner Belt & 18.4 & 57 & 0.76^{+0.13}_{-0.10} & 18.28^{+0.09}_{-0.09} \\
Intermediate Belt & 17.7 & 173 & 0.79^{+0.06}_{-0.06} & 16.96^{+0.1}_{-0.07} \\
Outer Belt & 16.56 & 108 & 0.86^{+0.12}_{-0.08} & 16.14^{+0.09}_{-0.06} \\
\enddata
\tablenotetext{a}{As defined in Section \ref{sec:orbfit}.}
\tablenotetext{b}{Limiting $H_{g'}$ magnitude for fitting.}
\tablenotetext{c}{Number of bodies brighter than $H_{lim}$.}
\tablenotetext{d}{Values are the median with a confidence interval of $\pm34\%$ from the distribution.}
\end{deluxetable}

\subsection{Color} \label{sec:color}
\begin{figure}
   \centering
   \includegraphics[width=\hsize]{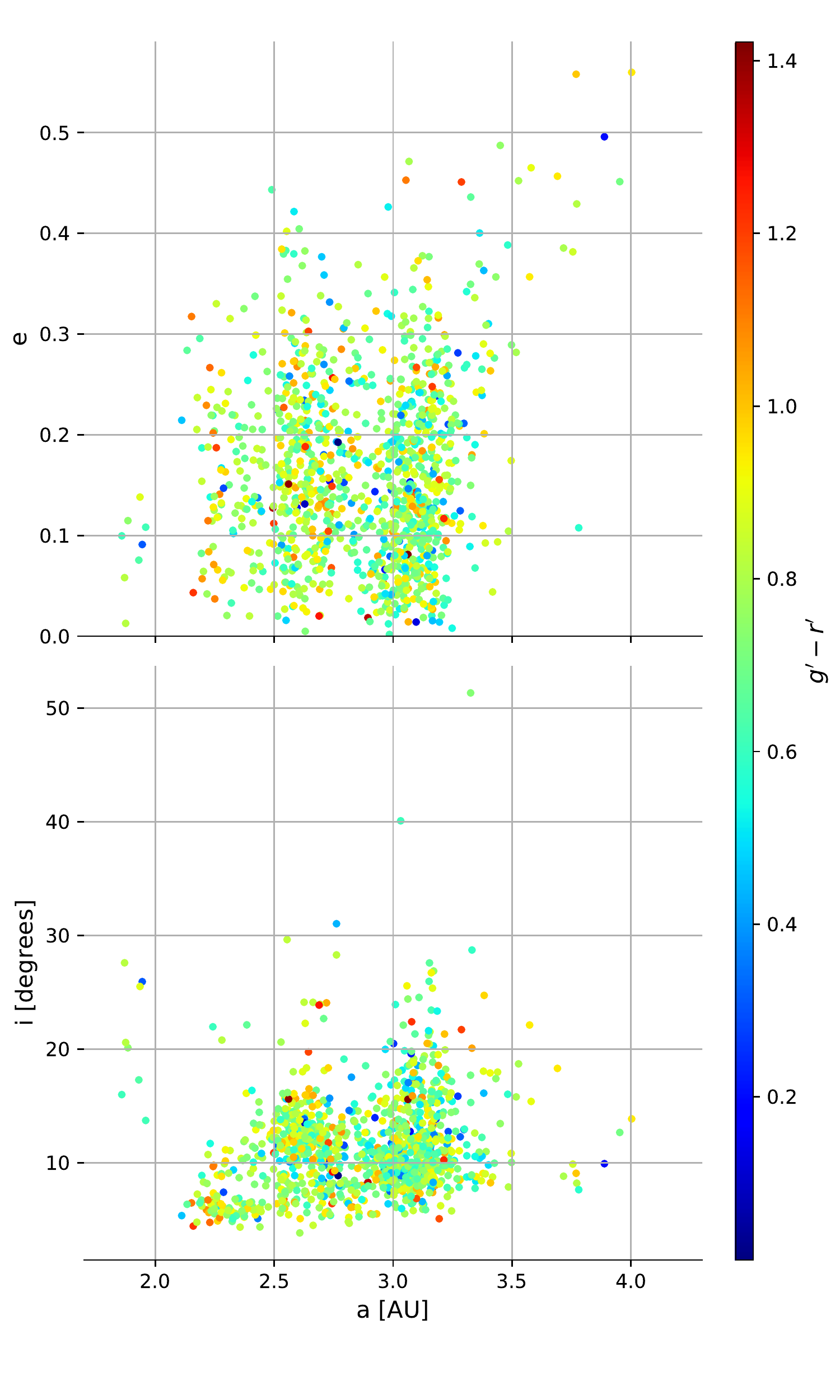}
   \caption{Same as Figure \ref{fig:orbits} but for Main Belt objects only that were also measured in $r$ with their respective $g'-r'$ color (measured from their $H_{g'}$ and $H_{r'}$ values). }
\label{fig:colors_aei}%
\end{figure}

\begin{figure}
   \centering
   \includegraphics[width=\hsize]{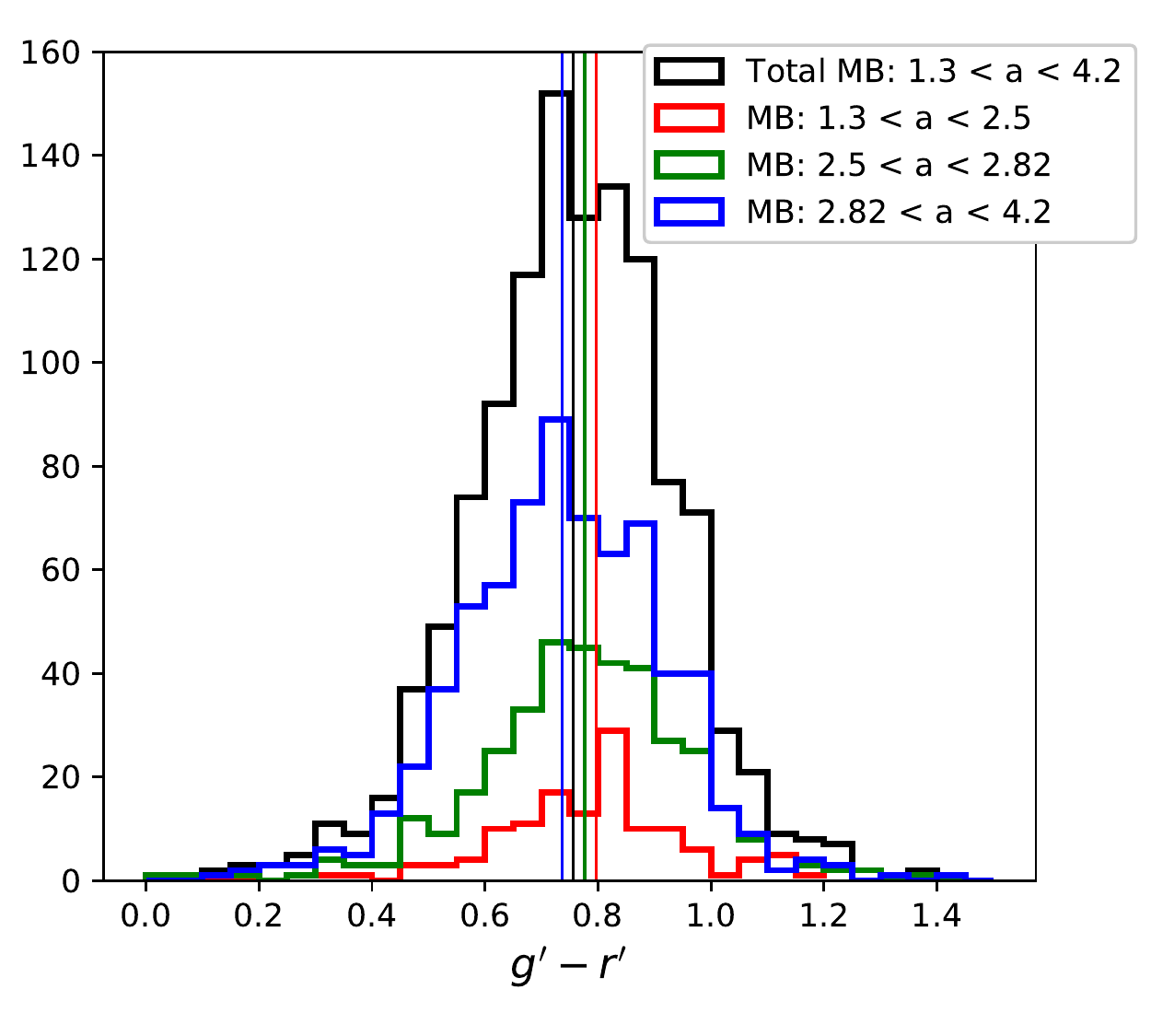}
   \caption{Histograms of the color of Main Belt asteroids. In black: all Main Belt asteroids; in red: Inner Belt; in green: Intermediate Belt; and in blue: Outer Main asteroids. Vertical lines mark the mean color for each population (see Table \ref{tab:colors}). Notice that the average $g'-r'$ colors for C and S asteroids (based in \citealt{Ivezic.2001}) are 0.45 and 0.65 respectively.}
\label{fig:color_hits}%
\end{figure}

The 2015A HiTS campaign observed in both $g'$ and $r'$, but most revisits were in $g'$, enabling us to detect tracks in both filters and produce colors for some of our tracks. We report $g'-r'$ colors from $H_{g'}-H_{r'}$ (section \ref{sec:H}) for the 1203 %bound and well constrained 
\textit{good tracks} measured in both bands. In Figure \ref{fig:colors_aei} we show these colors for the 1182 located in the region of the Main Belt as a function of their orbital parameters %(computed here)
and in Figure \ref{fig:color_hits} the $g'-r'$ distribution in the entire Main Belt is shown separated by class: Inner, Intermediate and Outer Belt. In both figures we observe that asteroids are mainly red, with a mean color of $g'-r'\sim0.756\pm0.008$ for all of them. Although we could not recognize any color--size dependency (section \ref{sec:H}), we recovered the known color-distance relationship \citep{Yoshida.2007, Gladman.2009}, as seen in Figure \ref{fig:color_hits}. Summarized in Table \ref{tab:colors} we show that MB asteroids get bluer as they get farther from the Sun. This dependency is usually explained as an asteroid type dependency: the outer belt would be dominated by C-type asteroids (bluer) and the inner belt would be dominated by S-type (redder); (as seen in many color plots; e.g., \citealt{Ivezic.2001, Yoshida.2007, Gladman.2009}). If we use \cite{Ivezic.2001} as a reference, the $g'-r'$ limit between C-- and S--types would be around $\sim0.55$, meaning that the vast majority of our asteroids would be S-type in the three MB divisions. We were not able, however, to distinguish any clear bimodality as in \cite{Ivezic.2001} to distinguish between types.

We explain the lack of the expected bimodality in color taking into account the asteroids' intrinsic lightcurves due to rotation. Most asteroids exhibit some variation with periods that range from $\sim$2 hours to $\sim$2.5 days (remember our 1.6 hour cadence), changing their brightness by 0.1 to 1.2 magnitudes \citep{Polishook.2012, Waszczak.2015}. Another posible contaminating source in our sample are NEOs, which exhibit similar rotational periods \citep{Vaduvescu.2017}. Our reported colors were measured as the average $H_{g'}$ (several values) minus the average $H_{r'}$, the latter generally being only one value measured at $\sim$1.6 hours from the nearest $g'$ measure. This means that colors reported in this work have a big uncertainty due to asteroid's rotation. In comparison, SDSS colors are measured within $\sim5$ minutes \cite{Ivezic.2001}. WISE, for example, observed in 4 bands simultaneously \citep{Wright.2010} allowing the measurement of $p_V$ for bodies with accurate orbital parameters and obtaining a strong bimodality associated with composition \citep{Masiero.2011}. %In our case we only have color data with 1.6 hours out of phase, which only allowed us to recover the known trend from red to blue colors with distance.

\begin{deluxetable}{cCCC}%[b!]
\tablecaption{$g'-r'$ colors for the Main Belt \label{tab:colors}}
\tablecolumns{4}
%\tablenum{1}
\tablewidth{0pt}
\tablehead{
\colhead{Population\tablenotemark{a}} &
\colhead{N\tablenotemark{b}} &
\colhead{Mean} &
\colhead{Stand. Dev.} %\\
%\colhead{(YYYY-mm-dd)} & \colhead{(d)} &
%\colhead{(arcsec)} & \colhead{} & \colhead{}
}
\startdata
Main Belt & 1182 & 0.756\pm.008 & 0.181\pm.008 \\ \hline
Inner Belt & 133 & 0.797\pm.031 & 0.175\pm.033 \\
Intermediate Belt & 369 & 0.776\pm.015 & 0.186\pm.015 \\
Outer Belt & 680 & 0.737\pm.010 & 0.177\pm.010
\enddata
\tablenotetext{a}{As defined in Section \ref{sec:orbfit}.}
\tablenotetext{b}{Number of bodies in each population.}
\tablecomments{The errors in the last two columns are computed taking into account the magnitudes errors and the confidence intervals shown in Section \ref{sec:orbfit} (Figure \ref{fig:orberrors}).}
\end{deluxetable}

Outside the Main Belt, we obtained colors for 12 known objects: %(we used their orbital parameters as delivered by JPL).
5 NEOs, 3 TNOs and 4 Jupiter family comets (JFCs). Colors for these bodies can be seen in Table \ref{tab:notMB}. NEOs have colors somewhat bluer than the MB's average color, which is consistent with \cite{Dandy.2003}, who claim the MB as a possible source, finding NEOs bluer than expected. The 4 JFCs are redder than the mean color of the MB, but inside the MB color range as seen in \cite{Solontoi.2012} (in $g'-r'$, with \citealt{Ivezic.2001} and this work for comparison) or in \cite{Lamy.2009} and  \cite{Jewitt.2015} (in $B-R$, with \citealt{Yoshida.2007} for comparison). For the TNOs, following the classification algorithm defined by \cite{Gladman.2008} and using the limit for scattered objects by \cite{Lykawka.2007}, (531017) 2012 BA$_{155}$ is a 2:5 resonant body, 2014 XW$_{40}$ is a scattered TNO and (523671) 2013 FZ$_{27}$ is in the limit between scattered and hot, outer classical TNO (and is also near the detached TNO zone). These 3 TNOs have red colors compatible with their respective families \citep{Sheppard.2010, Jewitt.2015, Pike.2017, Terai.2018}.

\begin{deluxetable}{lcCC}%[b!]
\tablecaption{$g'-r'$ colors for Non Main Belt known bodies \label{tab:notMB}}
\tablecolumns{4}
%\tablenum{1}
\tablewidth{0pt}
\tablehead{
\colhead{Name\tablenotemark{a}} & 
\colhead{Type}\tablenotemark{b} & 
\colhead{$g'-r'$}\tablenotemark{c} & %\\
\colhead{$H_{g'}$}
}
\startdata
2003 HU42 & NEO & 0.67\pm0.20 & 18.00\pm0.01 \\
2008 VU4 & NEO & 0.75\pm0.18 & 18.04\pm0.05 \\
2003 SS214 & NEO & 0.60\pm0.16 & 20.14\pm0.06 \\
2014 WL368 & NEO & 0.85\pm0.05 & 20.19\pm0.04 \\
2017 JB & NEO & 0.47\pm0.12 & 23.66\pm0.08 \\
%C/2015 D2 (PANSTARRS) & JFC &  0.80\pm0.04 & 13.29\pm0.02 \\
%P/2011 U2 (Bressi) & JFC & 1.064\pm0.030 & 13.96\pm0.01 \\
%C/2014 A5 (PANSTARRS) & JFC & 0.87\pm0.10 & 15.43\pm0.04 \\
C/2015 D2 & JFC &  0.80\pm0.04 & 13.29\pm0.02 \\
P/2011 U2 & JFC & 1.064\pm0.030 & 13.96\pm0.01 \\
C/2014 A5 & JFC & 0.87\pm0.10 & 15.43\pm0.04 \\
317P/WISE & JFC & 0.79\pm0.08 & 18.93\pm0.07 \\
2013 FZ27 & TNO & 0.92\pm0.05 & 4.72\pm0.02 \\
2012 BA155 & TNO & 1.17\pm0.15 & 6.57\pm0.10 \\
2014 XW40 & TNO & 1.25\pm0.13 & 6.79\pm0.07 \\
\enddata
\tablenotetext{a}{As identified by the MPC and JPL}
\tablenotetext{b}{NEO: Near Earth Object; JFC: Jupiter Family Comet (as identified by the JPL Small-Body Database Browser);
%\footnote{\url{https://ssd.jpl.nasa.gov/horizons.cgi}}); 
TNO: Trans-Neptunian Object}
\tablenotetext{c}{Colors computed as the difference of the averaged $H_{g'}$ and $H_{r'}$. The errors only consider magnitude uncertainty. To calculate $H$ we use orbital data ($r$ and $\Delta$) from JPL Horizons.}
%\tablecomments{}
\end{deluxetable}

\section{Summary and Discussion}\label{sec:conc}

Using data %associated with moving objects 
from the HiTS 2015 campaign, we found 5740 SS minor bodies. Considering only bodies with an observation arc longer than one night, we were able to identify 1738 MB asteroids (397 of them were known bodies), getting color information for 1182 of them.%asteroids.% from these MB bodies.

The luminosity function %cumulative distribution of the apparent magnitude was computed
for all bodies with apparent motions compatible with MB bodies (5703 in total) is well fit by a DPL, similar to the one found by \cite{Gladman.2009}, with a break in a similar magnitude but with much steeper slopes.

We found the %raw
size distribution for the MB population compatible with a SPL for the entire population as well as for the Inner, Intermediate and Outer MB separately. The slope parameters for these populations are much higher than previous surveys have reported, only comparable with values found by \cite{Parker.2008} at the bright end of the distribution.

%\cfg{ reconsider this paragraph, be more specific about the hypothesis that did not work, better yet say what might be missing from this analysis... albedo-size, family, etc}
%We simulated the MB populations based on our CSD results varying slope parameters and total number for each population but we were unable to replicate the apparent magnitude distribution.

We did not find a color--size dependence between 14 $<H_{g'}<$ 18 ($1<D<10$ km), as previously reported by \cite{August.2013} for similar sizes (analyzing over 7,000 bodies). \cite{Ivezic.2001} could not find a color--size relationship for similar objects analyzing $\sim$670,000 bodies in multiple filters.

\begin{figure}
   \centering
   \includegraphics[width=\hsize]{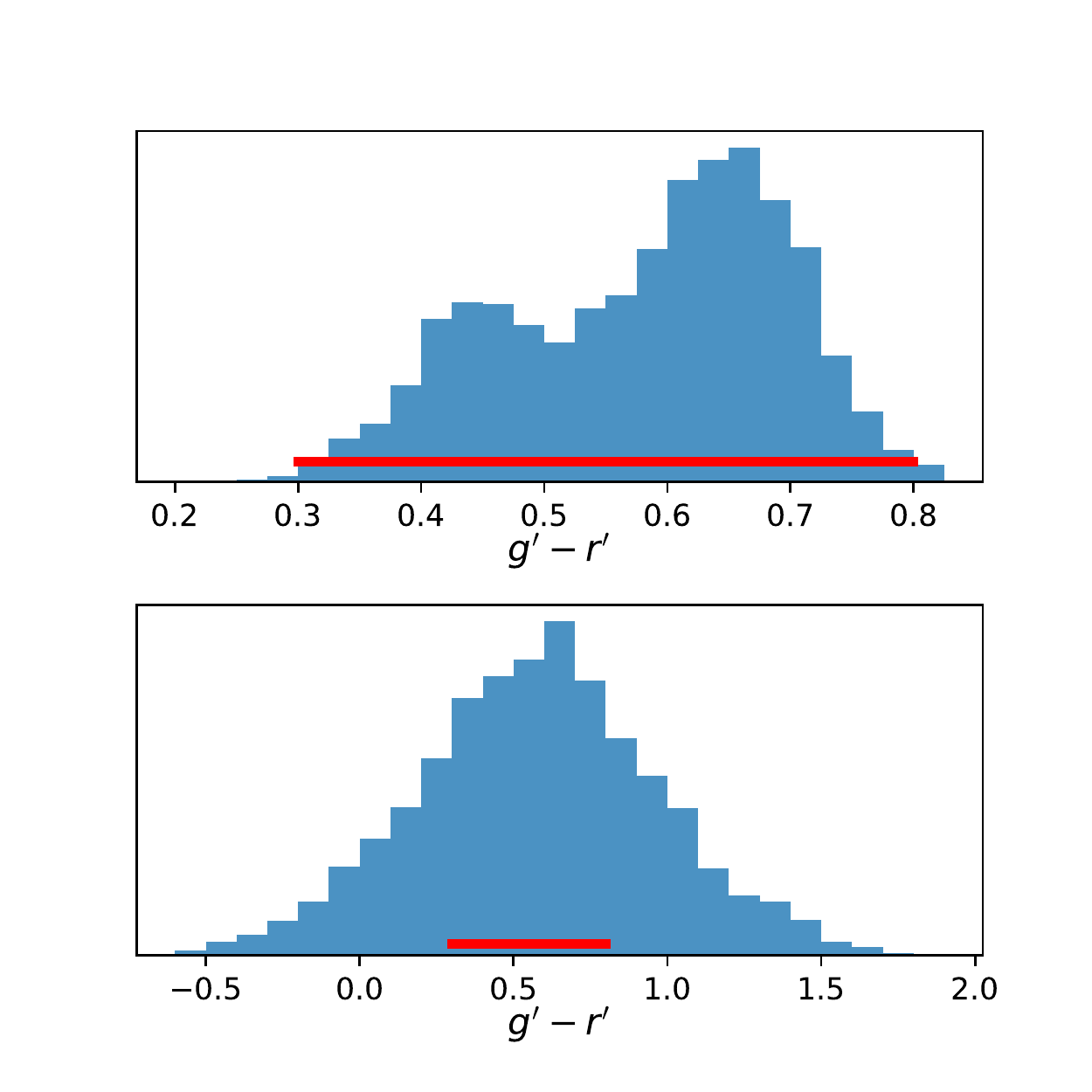}
   \caption{Top panel: simulated color bimodality resembling the one seen in \cite{Ivezic.2001}. Lower panel: apparent color due asteroids' rotation (see text). In both panels, the red line covers the same range, showing how the apparent color distribution gets wider because asteroids' rotation. }
\label{fig:sim_color}%
\end{figure}

\begin{figure}
   \centering
   \includegraphics[width=\hsize]{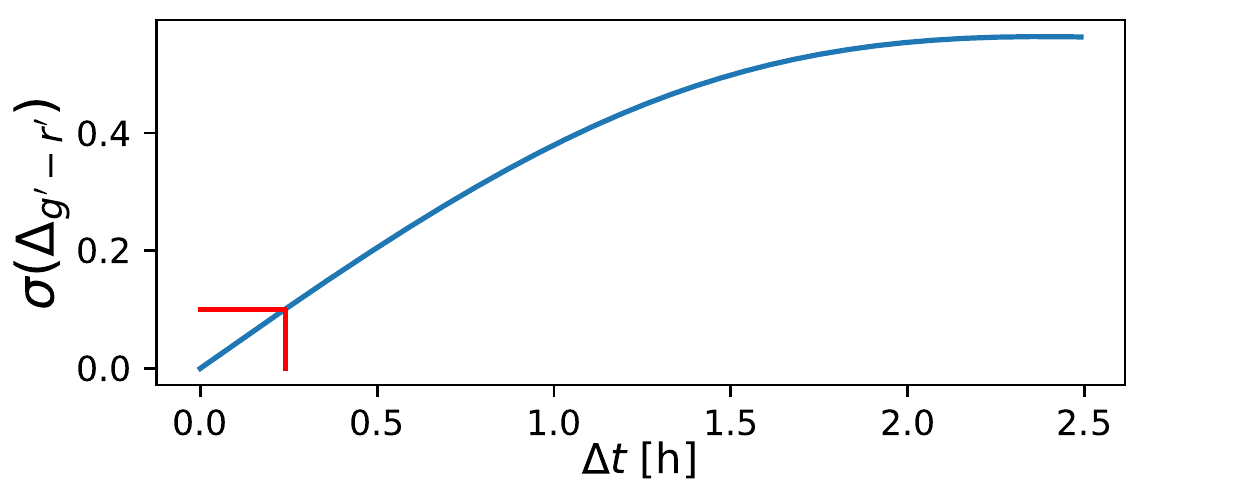}
   \caption{Standard deviation of the errors in color using consecutive measurements $\sigma(\Delta_{g'-r'})$ of the simulated population (see Figure \ref{fig:sim_color}) vs. time between observations $\Delta t$. In red we indicate the time separation $\Delta t = 0.24$hr that yields an uncertainty $\sigma(\Delta_{g'-r'})=0.1$ small enough to detect bimodality in the color distribution (see text).}
\label{fig:sim_dt}%
\end{figure}

The colors we report are most similar to S-type bodies in the MB, which is compatible with the scenario of the Outer MB (the most populous region of the MB) is dominated by S-type asteroids. We could not find any bimodality in color. In order to explain this we simulated an intrinsic bimodal population %in color
like the one by \cite{Ivezic.2001} (top panel of Figure \ref{fig:sim_color}). To measure how asteroids' rotation affects the measured color, we modeled their magnitudes as $m=m_0+A\sin(2\pi (\phi + f\Delta t))$ with $m$ the measured value, $m_0$ the mean magnitude (pseudo-randomly generated with equation \ref{eq:dpl} as a probability distribution), $A$ the magnitude variation (obtained from a pseudo-random triangular distribution between 0 and 1.2 magnitudes with the mode at 0.2), $f$ the rotation's frequency (obtained from a pseudo-random triangular distribution between 0.5 and 10 days$^{-1}$ with the mode at 0.5), $\phi$ the phase of the first observation (obtained from a pseudo-random flat distribution) and $\Delta t$ the time between one observation and another ($\Delta t=0$ for the simulated first observation and $\Delta t=1.6$ hours for the second observation in the HiTS 2015 case). The $A$ and $f$ distributions are based on \cite{Waszczak.2015}. Finally, the apparent color $g'-r'$ we obtained (with $g'$ obtained as $m_0$ and $r'$ as $m(\Delta t = 1.6 h)$) is shown in the lower panel of Figure \ref{fig:sim_color}, where the color distribution has broadened and lost any bimodality. We further estimated the effect of a time lag between filters on colors due to an asteroid's rotation. %showing that surveys aiming to measure colors need observations very near in time (as SDSS) or many observations in each band to get a proxy of $m_0$ in each band (in HiTS, we have a mean color for $g$, but for $r$ we have only 1 observation).
In Figure \ref{fig:sim_dt} we show the standard deviation of the errors between consecutive measurements of our simulated population ($\sigma(\Delta_{g'-r'})$) as a function of the time between those observations; since the two modes of our \textit{``true''} color distribution (top panel in Figure \ref{fig:sim_color}) are 0.2 magnitudes apart,
%and their standard deviation is 0.065
we set a rough $\sigma(\Delta_{g'-r'})$ limit of at least 0.1 to detect the bimodality. This value is found at $\Delta t=0.24$ hours or $\sim14$ minutes (red lines in Figure \ref{fig:sim_dt}). This is an important constraint for future surveys such as LSST \citep{LSST.2009}, which plans to visit one sky field twice a night, taking two images per visit. Since visits can be a few hours apart, ideally the two images per visit should be in different filters to retrieve good colors. Although LSST will deliver large lightcurves for many asteroids, allowing to measure colors even if the two images per visit were in the same filter \citep{Ivezic.2018}, this will be possible only for asteroids well tracked through many nights for a long time, while if the color is measured per visit we would be able to have colors for nightly tracklets even for those cases when the body is detected only once (generally the case for small bodies at the limiting magnitude).

%Another very interesting results are t
The slopes we got in our size distributions are much steeper than in any other survey, only comparable to the ones found by \cite{Parker.2008} for bright bodies. Our lower surface density of detections is consistent with \cite{Gladman.2009} if we take into consideration the amount of asteroids by ecliptic latitude found by \cite{Ryan.2009}, 
% as reference of observations at ecliptic latitude $0^{\circ}$ along with the ecliptic distribution found by
%and \cite{Ryan.2009}. %it seems we actually found the expected amount of asteroids, 
so our steep values would be caused by a lack of bright bodies (less numerous but able to flatten the distribution).
The apparent lack of bright bodies could be an effect of the ecliptic latitudinal distribution of asteroids on the observed luminosity function or it could be an effect of the analysis of HiTS moving objects. %or it could be a real effect of the MB morphology in the area and time surveyed. 
%If it is a defect in HiTS moving object analysis, 
The \textit{deep learning} analysis that distinguishes between real and bogus detections was designed for static (not elongated) transients and not specifically for moving objects, but there is no reason to believe it would discriminate bright bodies worst than faint bodies.%, although is unclear why this would affect only bright bodies.
%If our result is an effect of the actual MB morphology, it would be a sign of a lack of bright bodies for observations farther from the ecliptic.

\begin{figure}
   \centering
   \includegraphics[width=\hsize]{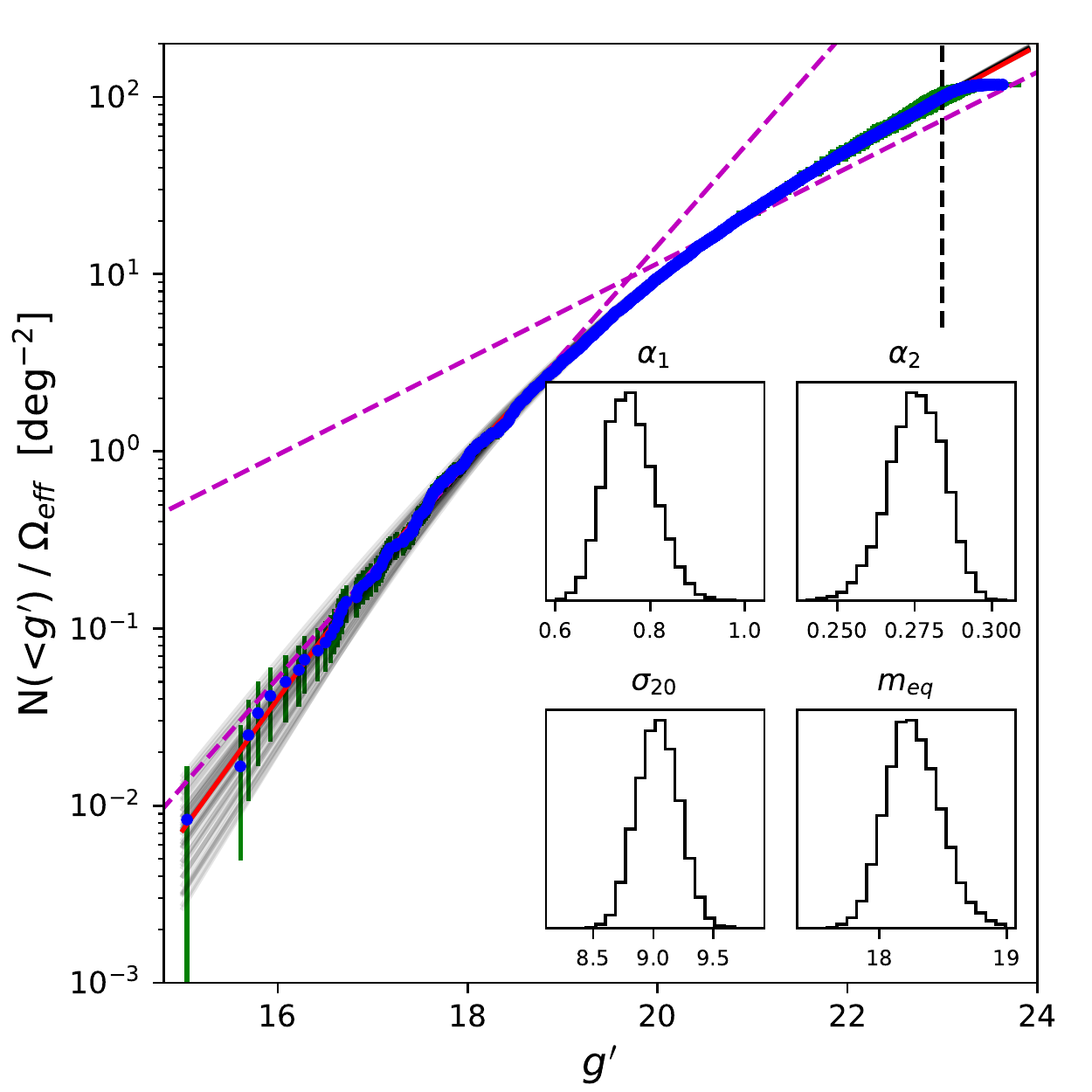}
   \caption{Cumulative distribution of magnitude $g'$ for the 12,928 tracks with MB-like velocities from the 2014 HiTS campaign (see section \ref{sec:conc}). A DPL was fit using MCMC to the 11,419 bodies brighter than 23 $g'$ (marked with a black dashed line). The small panels show the DPL parameter distributions. Using the mode with a $\pm$34\% confidence interval, we got $\alpha_1 = 0.76^{+0.05}_{-0.06}$, $\alpha_2 = 0.28^{+0.01}_{-0.01}$, $\sigma_{20} = 9.03^{+0.18}_{-0.18}$ and $m_{eq} = 18.16^{+0.33}_{-0.11}$. The red line shows the DPL given by the mode values and the gray lines show 50 random models from the MCMC procedure. In magenta dashed lines there is a proxy of the distribution by \cite{Gladman.2009}, considering they found $\sim100$ bodies at $R\sim 19.1$ at the bright end (slope of 0.61), $\sim1000$ bodies at $R\sim 23$ at the faint end (slope of 0.27), $\Omega= 8.4$deg$^2$ and approximating $g-R\sim 0.76$ (from Section \ref{sec:color}).}
\label{fig:cumg14}
\end{figure}

\begin{figure}
   \centering
   \includegraphics[width=\hsize]{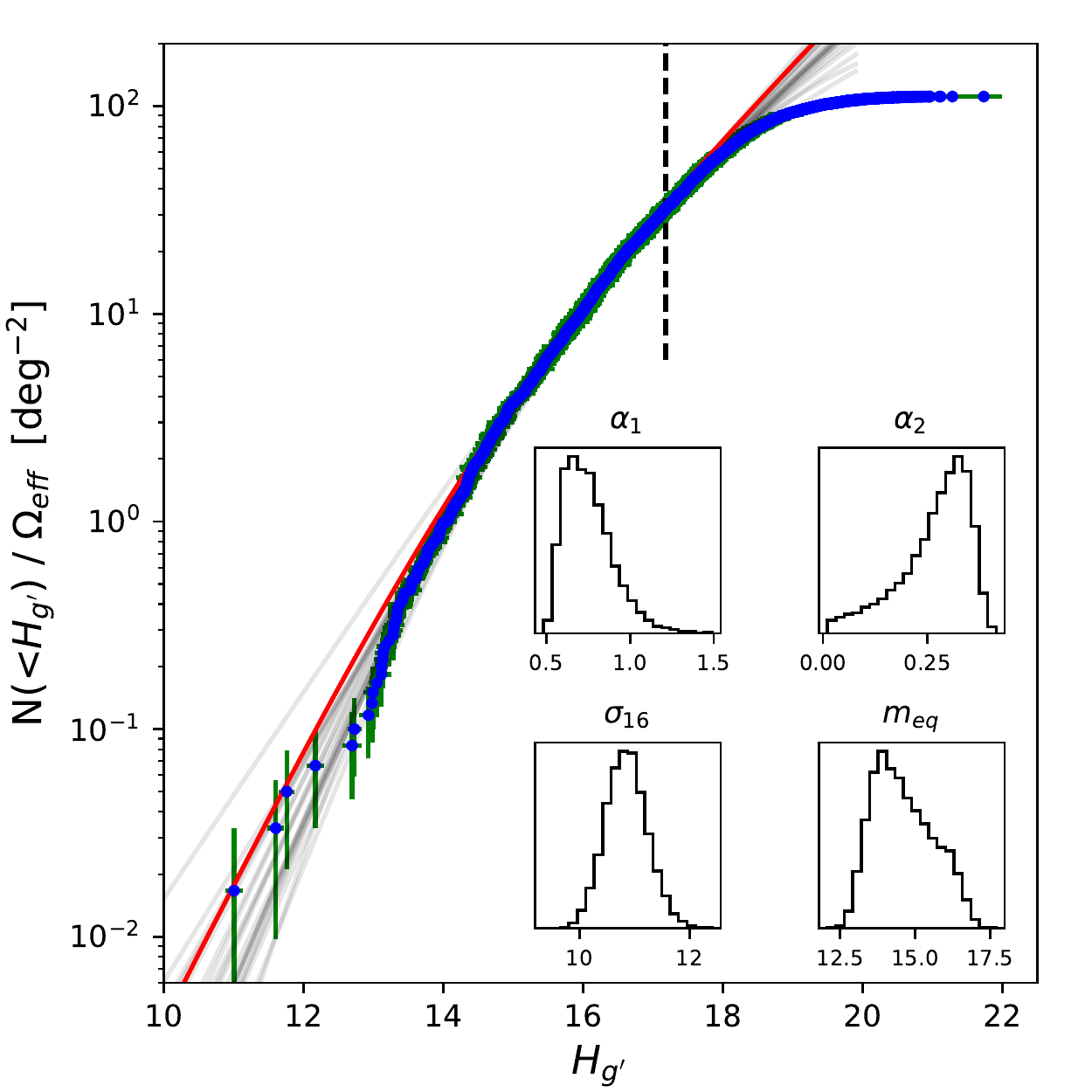}
   \caption{Cumulative $H$ distribution for all 6679 MB bodies  with good orbital solutions from the 2014 HiTS campaign (following the same criteria as for the 2015 HiTS campaign). A DPL (see text) was fit using MCMC to the 1729 bodies brighter than $H_{g'}=$ 17.06 (marked with a dashed line). The small panels show the DPL parameter distributions. Using the mode with a $\pm$34\% confidence interval, we got $\alpha_1 = 0.68^{+0.17}_{-0.09}$, $\alpha_2 = 0.34^{+0.04}_{-0.11}$, $\sigma_{16} = 10.85^{+0.40}_{-0.40}$ and $m_{eq} = 13.82^{+1.63}_{-0.51}$. The red line shows the DPL given by the mode values and the gray lines show 20 random models from the MCMC procedure.}
\label{fig:cumH14}
\end{figure}

%To assess if the steep slopes are an effect of the MB ecliptic distribution 
We explored the latitudinal dependence by analyzing the SD of asteroids from the HiTS 2014 campaign \citep{Pena.2018}. Those asteroids were found mainly between ecliptic latitudes $0^{\circ}$ and $15^{\circ}$. Using bodies with an apparent ecliptic latitude velocity compatible with the MB (less than $-0.16^{\circ}$day$^{-1}$ instead of less than $-0.15^{\circ}$day$^{-1}$ as for the 2015 data because in 2014 campaign we have Jupiter trojans around $-0.15^{\circ}$day$^{-1}$) and following the same procedure as in Section \ref{sec:mag} (using $m_{50}=23.6$ and $\Delta m_{50}$=1.1 for $\eta$ in equation \ref{eq:eta} and $\Omega=120$deg$^2$ for the surveyed area) we obtained the distribution shown in Figure \ref{fig:cumg14}, which exhibits a slope of $\sim 0.76$ for the bright end and a slope of $\sim .28$ for the faint end. These values are very similar to (although somewhat steeper than) the values found by \citealt{Gladman.2009} (see dashed lines from Figure \ref{fig:cumg14}). Continuing the analysis of the 2014 data, we computed  the SD for all ``\textit{good}'' tracks (using the same criteria from Section \ref{sec:orbfit}) and fitted a DPL to it (an SPL was not enough to fit the data). This DPL has the same form as equation \ref{eq:dpl}, but instead of using magnitude 20 as reference, we used magnitude 16. The result is shown in Figure \ref{fig:cumH14}, where the DPL was fitted using data up to a limiting magnitude $H_{g'}=$ 17.18 (using a limiting magnitude of $g'=$ 22.5 for 90\% completeness and a phase angle of less than $9.2^{\circ}$). The resulting slopes are $0.68^{+0.17}_{-0.09}$ at the bright end and $0.34^{+0.04}_{-0.11}$ at the faint end, which is consistent with previous results such as \cite{Ivezic.2001} and \cite{Gladman.2009}. This supports the hypothesis that the steeper slopes for our 2015 campaign are due to a lack of bright sources %because of their scarcity that cause we don't see them 
at higher latitudes. %, while at latitudes near zero we see a superposition of bodies with all kind of inclinations, making easy to find those bright bodies. 
A similar result was found by \cite{Bhattacharya.2010}, where bodies at latitude $\sim 5^{\circ}$ have fluxes 30\% fainter than those at latitude $\sim 0^{\circ}$. 
%To compensate this ecliptic dependence of the cumulative distributions it would be necessary to account for the ecliptic dependence of the asteroids abundance, namely, add the latitude distribution derived by \cite{Brown.2001} to the analysis as in \citealt{Fuentes.2011} (although for the MB we do not have the width of the distribution necessary for this analysis).
We will further investigate the effect of this latitudinal distribution of asteroids on the observed luminosity function and the apparent lack of bright bodies in our survey.

HiTS as a high cadence survey was a precursor of wider future surveys such as the LSST. Although it was not optimized specifically for finding asteroids, its wide coverage and high cadence \sout{of HiTS} proved to be appropriate for discovering asteroids. We found that having at least three observations of the same field in a night (\textit{tracklet}) is important for this task. Fitting an accurate orbit for a new discovery requires observations in at least two nights, which for the sparse sky coverage of HiTS is more likely if the same field is visited in consecutive nights while objects are still there. LSST will visit the same sky $\sim3$ nights later, improving the orbit estimation but hindering the pairing of \textit{tracklets} between nights. Another important quality of HiTS is its machine learning vetting algorithm that discriminates true sources from bogus; this speeds up the linking of asteroids and diminishes the probability of mixing unrelated detections. Finally, HiTS cadence, although excellent for asteroids discovery, is not  well suited for color measurements due to asteroid rotation. Surveys should plan on taking consecutive multifilter observations for accurate color measurements whenever observing conditions are photometric.

\section{Acknowledgments}
J.P. acknowledges the support from CONICYT Chile through (CONICYT-PFCHA / Doctorado-Nacional / 2017-21171752). 
J.P., C.F., F.F., J.S.M., G.C.V., S.G.G. and J.M. acknowledge support from Grant PIA AFB-170001, Centro de Modelamiento Matem\'atico (CMM), Universidad de Chile.
F.F., M.H., S.G.G., P.A.E., G.C.V., and J.M. acknowledge support from the Ministry of Economy, Development, and Tourism’s Millennium Science Initiative through grant IC120009, awarded to The Millennium Institute of Astrophysics (MAS).
J.P., C.F. acknowledge support from the BASAL Centro de Astrof\'isica y Tecnolog\'ias Afines (CATA) PFB-06/2007.
F.F. acknowledges support from Conicyt through the Fondecyt Initiation into Research project No. 11130228.
S.G.G. and L.G. acknowledge support from FONDECYT postdoctoral grants 3130680 and 3140566, respectively.
G.C.V. acknowledges support from CONICYT through the FONDECYT Initiation grant No. 11191130
P.A.E. and P.H. acknowledge support from FONDECYT regular grants 1171678 and 1170305, respectively.
%F.F., J.C.M., P.H., G.C.V., and P.A.E. acknowledge support from Conicyt through the Programme of International Cooperation project. 
L.G. was funded by the European Union's Horizon 2020 research and innovation programme under the Marie Sk{\l}odowska--Curie grant agreement No. 839090.
S.G.G. acknowledges support of FCT under Project CRISP PTDC/FIS-AST- 31546.
J.M. acknowledges the support from CONICYT Chile through CONICYT-PFCHA/Doctorado-Nacional/2014-21140892. 
TdJ was funded by the Bengier Postdoctoral Fellowship and is grateful to Gary and Cynthia Bengier for their support.
Powered{@}NLHPC: this research was partially supported by the supercomputing infrastructure of the NLHPC (ECM-02).
Part of this work was done under the Harvard-Chile data science school.
This project used data obtained with the Dark Energy Camera (DECam), which was constructed by the Dark Energy Survey (DES) collaboration.
Funding for the DES Projects has been provided by the U.S. Department of Energy, the U.S. National Science Foundation, the Ministry of Science and Education of Spain, the Science and Technology Facilities Council of the United Kingdom, the Higher Education Funding Council for England, the National Center for Supercomputing Applications at the University of Illinois at Urbana--Champaign, the Kavli Institute of Cosmological Physics at the University of Chicago, Center for Cosmology and Astro--Particle Physics at the Ohio State University, the Mitchell Institute for Fundamental Physics and Astronomy at Texas A\&M University, Financiadora de Estudos e Projetos, Funda\c{c}\~ao Carlos Chagas Filho de Amparo, Financiadora de Estudos e Projetos, Funda\c{c}\~ao Carlos Chagas Filho de Amparo \`a Pesquisa do Estado do Rio de Janeiro, Conselho Nacional de Desenvolvimento Cient\'ifico e Tecnol\'ogico and the Minist\'erio da Ci\^encia, Tecnologia e Inova\c{c}\~ao, the Deutsche Forschungsgemeinschaft and the Collaborating Institutions in the Dark Energy Survey. 
The Collaborating Institutions are Argonne National Laboratory, the University of California at Santa Cruz, the University of Cambridge, Centro de Investigaciones Energ\'eticas, Medioambientales y Tecnol\'ogicas--Madrid, the University of Chicago, University College London, the DES--Brazil Consortium, the University of Edinburgh, the Eidgen\"ossische Technische Hochschule (ETH) Z\"urich, Fermi National Accelerator Laboratory, the University of Illinois at Urbana--Champaign, the Institut de Ci\`encies de l'Espai (IEEC/CSIC), the Institut de F\'isica d'Altes Energies, Lawrence Berkeley National Laboratory, the Ludwig--Maximilians Universit\"at M\"unchen and the associated Excellence Cluster Universe, the University of Michigan, the National Optical Astronomy Observatory, the University of Nottingham, the Ohio State University, the University of Pennsylvania, the University of Portsmouth, SLAC National Accelerator Laboratory, Stanford University, the University of Sussex, and Texas A\&M University.

\appendix
\section{Transforming Observer's Coordinates to Barycentric Coordinates}
\label{appendix:1}
Making \textit{tracklets} (collections of detections that resemble a linear trajectory) is easy if you consider detections of only one night, but to link detections from one night to another proves to be challenging (linear fitting does not work all the times and we do not always find a body every night). To solve this, we decided to  change our coordinate reference to the barycenter of our solar system, where the coordinates of our tracklets should resemble straight lines and linear extrapolation to join different bodies would easily work. The problem is that to do this we needed to know the distance of these bodies to the observer's position or to the barycenter. This meant that we had to assume different distances to look for linear trajectories in the barycentric frame.

Assuming that all detections are at distance $r$ to the barycenter, we needed to know the position of the Earth in the barycentric frame ($\mathbf{r_E}$) to solve all the geometry. To make the transformations from the observer's frame to the barycentric frame we used the module \texttt{SkyCoord} from the \texttt{astropy} package \citep{astropy.2013, astropy.2018} of \texttt{Python}. To do this, having the equatorial coordinates of the bodies as seen by the observer together with the observer's position we needed to measure the distance $\Delta$ between the observer and the bodies. Using trigonometry, it is easy to see that $\Delta$ is given by equation \ref{eq:a1delta} (solving the quadratic equation given by Equation (\ref{eq:a1costeo})\footnote{You could be tempted to replace $r_E^2\cos^2{\phi}-r_E^2$ by $r_E^2\sin^2{\phi}$ but if you are using \texttt{Python}'s \texttt{numpy} it will give you large errors for $\Delta$.}), where $\phi$ is the elongation (angle between the body and the barycenter)\footnote{\texttt{astropy} easily allows us to know the Sun's position seen by the observer, which is corrected by the light's travel time. That is \textit{not} the position we want, but the actual one at observing time. We managed to get this last one by using the observer's position delivered by \texttt{astropy.coordinates.EarthLocation} ($\mathbf{r_E}$) transformed to the barycentric frame and then inverting the sign of their cartesian values to get the Sun's position in the observer's frame.}.

\begin{eqnarray}
    \label{eq:a1costeo}
    r^2 &=& r_E^2 + \Delta^2 - 2r_E\Delta\cos{\phi} \\
    \label{eq:a1delta}
    \Delta &=& r_E\cos{\phi} + \sqrt{r_E^2\cos^2{\phi} - r_E^2 + r^2}
\end{eqnarray}

Having the equatorial coordinates from the observer, the observer's position $\mathbf{r_E}$ and the distance from the observer to the body $\Delta$, it is possible to move the origin of the body's coordinates to the barycenter. An example of the kind of equations you need for this are in \cite{Bernstein.2000}. To make this transformation, we used the simple interface facilitated by \texttt{python}'s \texttt{astropy.SkyCoord} (along with other \texttt{astropy}'s functionality such as ``\texttt{time}'', ``\texttt{units}'', etc). The frame we used is that of the barycentric ecliptic coordinates.

\section{Night--to--Night Linking Algorithm}
\label{appendix:2}
Once we had the barycentric ecliptic coordinates for all tracklets (assuming a body-observer distance $r$), we estimated their position at two different times using a linear fitting on their coordinates. This fit used the corrected time $t'$ approximated by $t' = t - 1/c$, where $t$ is the observed time for each coordinate and $c$ the speed of light.

For each of the estimated times, we performed a neighbor finding routine using a k-d tree algorithm\footnote{Implemented in \texttt{Python} using the \texttt{scikit-learn} package (\url{https://scikit-learn.org/}).} on their estimated coordinates to associate a tracklet to others if their estimated positions were close enough. To cluster one tracklet $tr_{j}$ to other tracklets ($\{tr_i\}_{i\neq j}$) we took care that in $\{tr_i\}_{i\neq j}$ there were no tracklets from the same night as $tr_j$. This did not stop $\{tr_i\}_{i\neq j}$ from having tracklets in the same night. To manage this problem, we divided the cluster $\{tr_j \cup \{tr_i\}_{i\neq j}\}$ into as many clusters as necessary so there were no repeated nights in any of the final clusters. Having done this clustering in both times independently, we crossed both sets of clusters to end with a collection where every clustered tracklets must have been clustered in both estimated times. We called any of this final clusters \emph{tracks}.

The times for the estimated positions were chosen to fall at roughly one and three quarters of the total arc of this survey (57,073 and 57,077 in modified Julian dates). The radius to make the clustering was based on the necessary radius to cluster most of the known bodies that were found among the tracklets without joining tracklets that did not correspond to the known asteroids. The radius that optimized both criteria was found to be 0.01 degrees. Using other radii (namely, getting more tracks in only one night or mixing tracklets that were not the same body) increased the error when estimating orbital parameters (see section \ref{sec:orbfit}).

Since we split clusters so they all were from different nights we could end up with tracklets repeated in more than one cluster. For these cases, we repeated the coordinate estimation process but used estimation times falling at 1/4 and 3/4 of the time arc of the cluster and we measured the maximum distance between estimated positions of the tracklets in the cluster (namely, the \emph{cluster error}). Finally we left the repeated tracklet in the cluster that had the smallest \emph{cluster error} and we removed it from the other clusters it belonged to.

%\clearpage

%%%%\clearpage

\end{document}